\documentclass[
aps,
reprint,
amsmath,amssymb,prl
]{revtex4-2}

\usepackage{blindtext}
\usepackage{graphicx}% Include figure files
\graphicspath{{./Figures}}
\usepackage{dcolumn} % Align table columns on decimal point
\usepackage{bm} % bold math
\usepackage{braket}
\usepackage{xcolor}
\usepackage{dsfont}
\usepackage{comment}
\usepackage{tikz}
\usepackage{zx-calculus}
\usepackage{booktabs}
\usepackage{amsmath}
\usepackage{subcaption}
\usepackage{wrapfig}
\usepackage[colorlinks=true, linkcolor=blue, citecolor=blue, urlcolor=magenta]{hyperref}
\usepackage{xparse}

\NewDocumentCommand{\tens}{e{_^}}{%
  \mathbin{\mathop{\otimes}\displaylimits
    \IfValueT{#1}{_{#1}}
    \IfValueT{#2}{^{#2}}
  }%
}

\def \beq {\begin{eqnarray}}
\def \eeq {\end{eqnarray}}

\begin{document}

\title{Unveiling a Hidden Percolation Transition in Monitored Clifford Circuits:\\ Inroads from ZX-Calculus}
\date{February 2024}

\author{Einat Buznach Ahituv}
\affiliation{Department of Physics, Bar-Ilan University, 52900, Ramat Gan, Israel}

\author{Debanjan Chowdhury}
\affiliation{Department of Physics, Cornell University, Ithaca, New York 14853, USA}

\author{Jonathan Ruhman}
\affiliation{Department of Physics, Bar-Ilan University, 52900, Ramat Gan, Israel}

\begin{abstract}
We revisit the measurement-induced phase transition (MPT) in Clifford circuits, which are both classically simulable and exhibit critical behavior widely believed to be distinct from classical percolation theory, using ZX-calculus. We analyze the MPT in a dynamical model composed of CNOT, SWAP, identity gates, and Bell-pair measurements, respectively, arranged randomly in a brickwork pattern. Our circuits exhibit a transition that is seemingly distinct from classical percolation based on standard arguments, that is in line with the prevailing understanding in the field. 
In contrast, by employing ZX-calculus based simplification techniques, we unveil a hidden percolation transition within the circuit structure. We demonstrate that the classical percolation transition in the ZX-simplified network coincides with the MPT observed through mutual information. Our findings suggest that the MPT in Clifford circuits is, in fact, controlled by a classical percolation transition in disguise.
\end{abstract}

\maketitle

{\it Introduction.-} Measurement-induced phase transitions (MPTs) represent a class of dynamical transitions in quantum systems subject to unitary evolution, interspersed with measurements \cite{skinner2019measurement,li2018quantum,chan2019unitary}. These transitions have been recently realized on quantum simulators~\cite{Noel2022,koh2023measurement,google2023measurement} and are important to a number of fields, including open quantum systems, non-equilibrium phenomena, quantum computation, and quantum error correction \cite{choi2020quantum,gullans2020dynamical,fan2021self,li2021statistical,lovas2024quantum}. Namely, for models with a finite local Hilbert space dimension ($q$), the competing interplay between the unitary evolution and measurements drives the system from an ``entangling" phase to a ``disentangling" phase as the measurement rate, $p$, exceeds a critical threshold, $p_q$; see Refs.~\cite{gullans2020dynamical,jonay2018coarsegraineddynamicsoperatorstate,nahum2021measurement}. A key question surrounding MPTs, that has been at the center of much recent attention, concerns their critical theory and universality class. 

Certain limiting cases have enjoyed remarkable theoretical progress. Free-fermion models exhibit a diverse range of phases and transitions, with a well-established field theory description \cite{nahum2020entanglement,chahine2024entanglement,KlockeMajoranaLoops,fava2023nonlinear,jian2022criticality,leung2023theory,poboiko2024measurement}; recently this analysis was extended to interacting fermions \cite{poboiko2024measurement2,Guo2024measurement}. Other tractable examples involve the limit  $q\rightarrow\infty$ \cite{vasseur2019entanglement,bao2020theory}, where the problem can be mapped to classical spin models~\cite{hayden2016holographic,zhou2019emergent} and the MPT is described by a classical percolation transition~\cite{jian2020measurement,bao2020theory}.  The connection to percolation in the $q\to\infty$ limit follows from the ``minimal-cut picture"~\cite{nahum2017quantum}, where the minimal length of a cut in the spacetime circuit determines the R\'enyi entropies, which are all equal in Haar random unitary circuits at $q\to\infty$.   In this case the MPT marks the point where the circuit becomes physically disconnected~\cite{skinner2019measurement}.

A complete theoretical understanding for a finite $q$ and for generic systems remains elusive. 
The Hartley function (the R\'enyi entropy of degree zero), which is not affected by $q$, still undergoes a classical percolation transition at $p=p_c$, marking the  physical disconnection point. However, the MPT for non-zero R\'enyi entropy occurs at a smaller value at $p = p_q<p_c$ within a connected circuit~\cite{skinner2019measurement,zabalo2020critical}. 
With increasing $q$, these transitions merge, with the MPT ultimately governed by classical percolation for $q\rightarrow\infty$. 
Recent studies have revealed significant deviations in the universality class of the MPT from the nearby percolation transition~\cite{zabalo2022operator,kumar2024boundary}; on quantum trees, MPT exhibits Berezinskii-Kosterlitz-Thouless (BKT)-like scaling~\cite{nahum2021measurement}. While numerical investigations on small systems with careful mitigation of finite-size effects provide valuable insights, accurately capturing the universal behavior remains challenging.

Clifford circuits allow for investigation of large system sizes, and serve as a useful platform for studying MPT~\cite{li2018quantum,li2019measurement,li2021conformal,gullans2020dynamical,ippoliti2021entanglement,sang2021measurement,vijay2020measurement,lavasani2021measurement}. Indeed, Clifford circuits exhibit an MPT at a point distinct from the classical percolation transition, and with distinct critical properties~\cite{zabalo2020critical,zabalo2022operator,kumar2024boundary}.  The minimal-cut method to quantify entanglement, which generically leads to percolation as noted previously, overestimates the entanglement in this case due to the limited and non-universal set of gates. For example as we will see, the CNOT gate, which is the entangling gate, can reduce the entanglement rather than enhance it with finite probability.  At the same time,  under Clifford dynamics the entanglement spectrum  is featureless, such that all R\'enyi entropies undergo the transition at the same point, including the Hartley function.

This raises the important question of whether the MPT on Clifford circuits represents a truly novel universality class, or if a percolation-type transition emerges in terms of an {\it a priori} unknown representation. In this work, we employ ZX-calculus~\cite{coecke2008interacting,coecke2018picturing,Duncan2020graphtheoretic,van2020zx} --- a graphical language for representing and simplifying quantum circuits --- to unveil a percolation transition in a class of random Clifford circuits undergoing MPT. We demonstrate that the classical percolation transition in ZX-simplified circuits coincides with the MPT identified via mutual information.  Our findings suggest that the universality class for the MPT in Clifford circuits is indeed governed by an underlying classical percolation phenomenon.

{\it Model.-} We consider a simplified model featuring $N$ qubits arranged in a linear one-dimensional chain, which captures all of the essential features tied to the MPT in Clifford circuits, with two-site unitary operations applied in a brickwork pattern. The initial state is chosen to be a tensor product of nearest neighbor Bell-pairs across odd bonds. However, the results are not sensitive to the initial condition \cite{supp}. We employ four distinct types of two-site operations: Control-NOT (CNOT) gates, SWAP gates, Identity gates, and finally, measurements in the Bell-pair basis. Each operation is selected randomly with a specific probability $r(1-p)$, $(1-r)(1-p)$, $p/2$ and $p/2$, respectively, as outlined in Table \ref{tab:probs}. For the CNOT gates, two configurations are considered: one with the left qubit acting as control and the other with the roles reversed. These configurations are chosen randomly with equal probability during circuit construction. We note that our dynamical model offers some unique advantages. While the circuit does not undergo a classical percolation transition for any value of $r$ and $p$, the model is sufficiently complex to undergo an  MPT and ideally suited for leveraging the simplifications enabled by ZX-calculus.

The dynamics of the model are efficiently simulated using the Gottesman-Knill theorem~\cite{gottesman1998,nest2008classical,biswas2024classical}. This approach involves evolving a set of $N$ stabilizers, $\{ \mathcal O_j \}_{j=1}^N$, Pauli operators satisfying the algebra, $[\mathcal O_i, \mathcal O_j] = \mathbb I \,\delta_{ij}$, that uniquely define the quantum state up to phases. Each stabilizer is represented by a binary vector of length $2N$. The application of Clifford gates, $U$, and projective measurements~\cite{li2018quantum}, maps these stabilizers to a new set of stabilizers, e.g. $\mathcal O_j' = U \mathcal O_j U^\dag$, where importantly, the number of stabilizers is conserved during time evolution. We can efficiently evolve the state (almost) indefinitely in time, and for large system sizes with polynomial efficiency.

\begin{table}
    \centering
    \begin{tabular}{c|cccc}
        Operation & CNOT & SWAP & Identity & Bell-pair \\\hline
        & & & &\\
        Probability & r(1-p) & (1-r)(1-p) & p/2 & p/2 \\
         & & & &\\
         Symbol  & \includegraphics[width=0.25in]{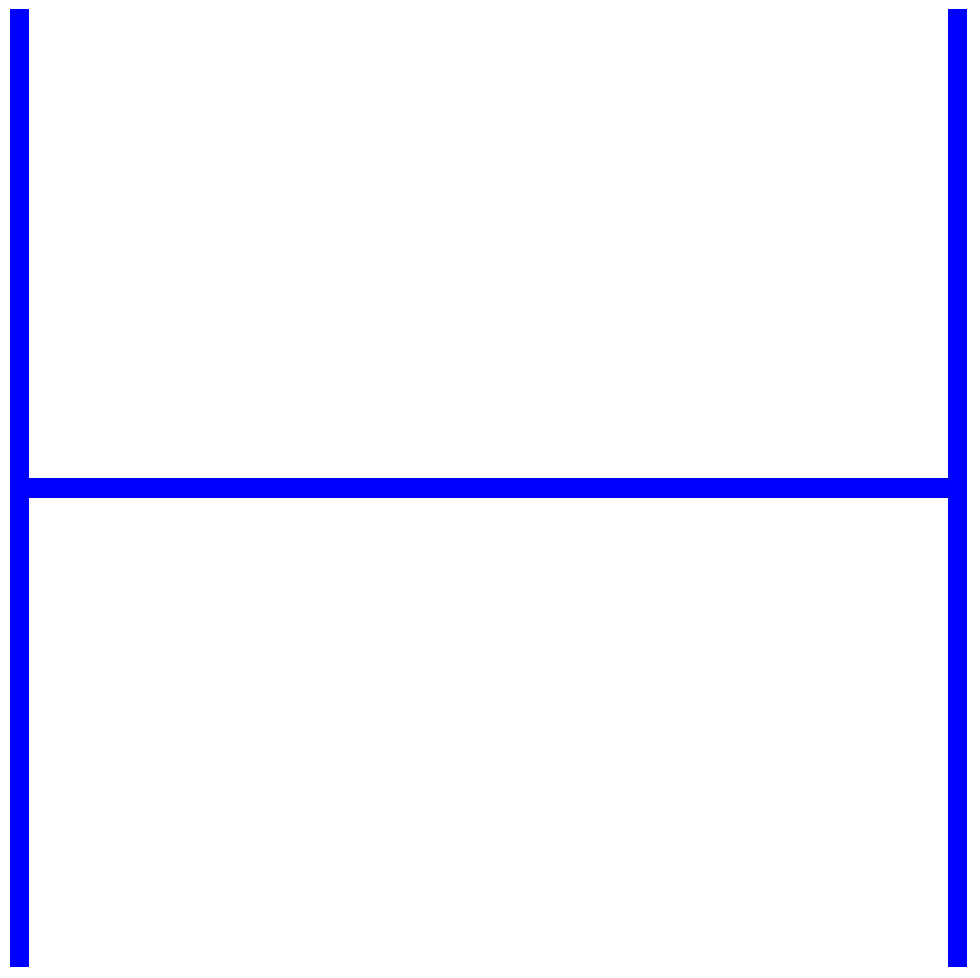} & \includegraphics[width=0.25in]{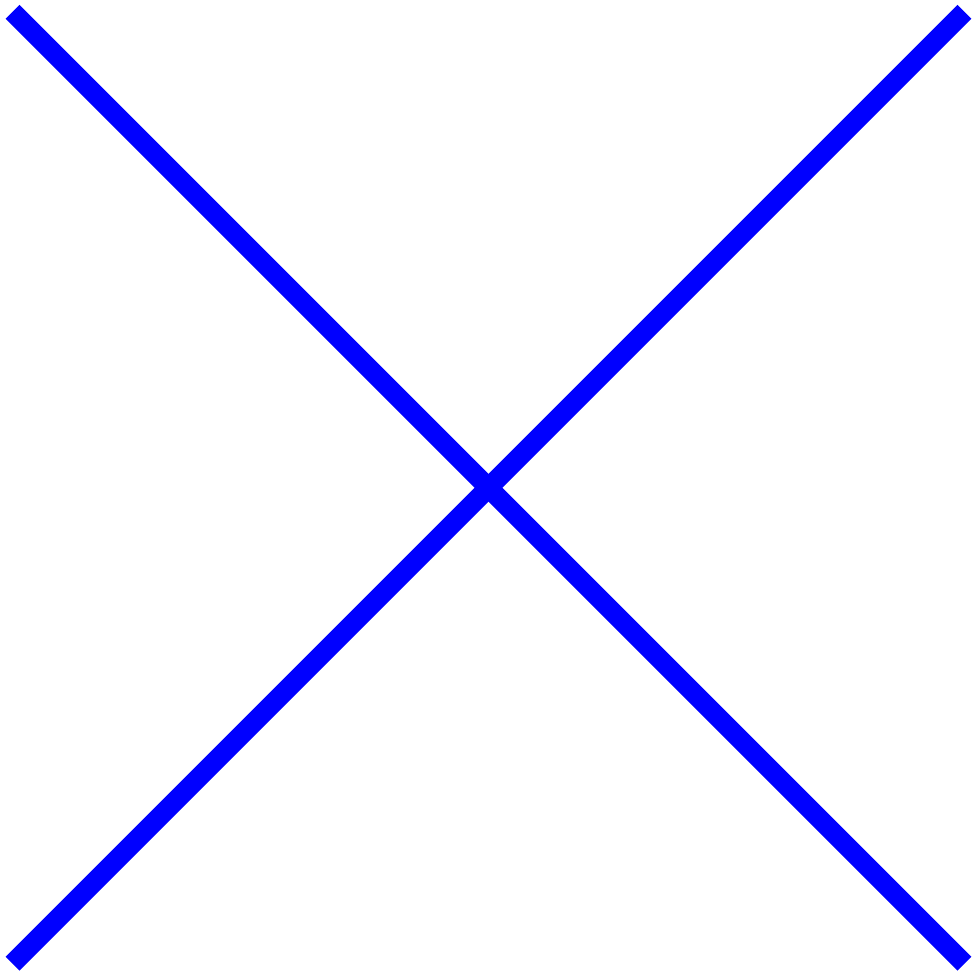} & \includegraphics[width=0.25in]{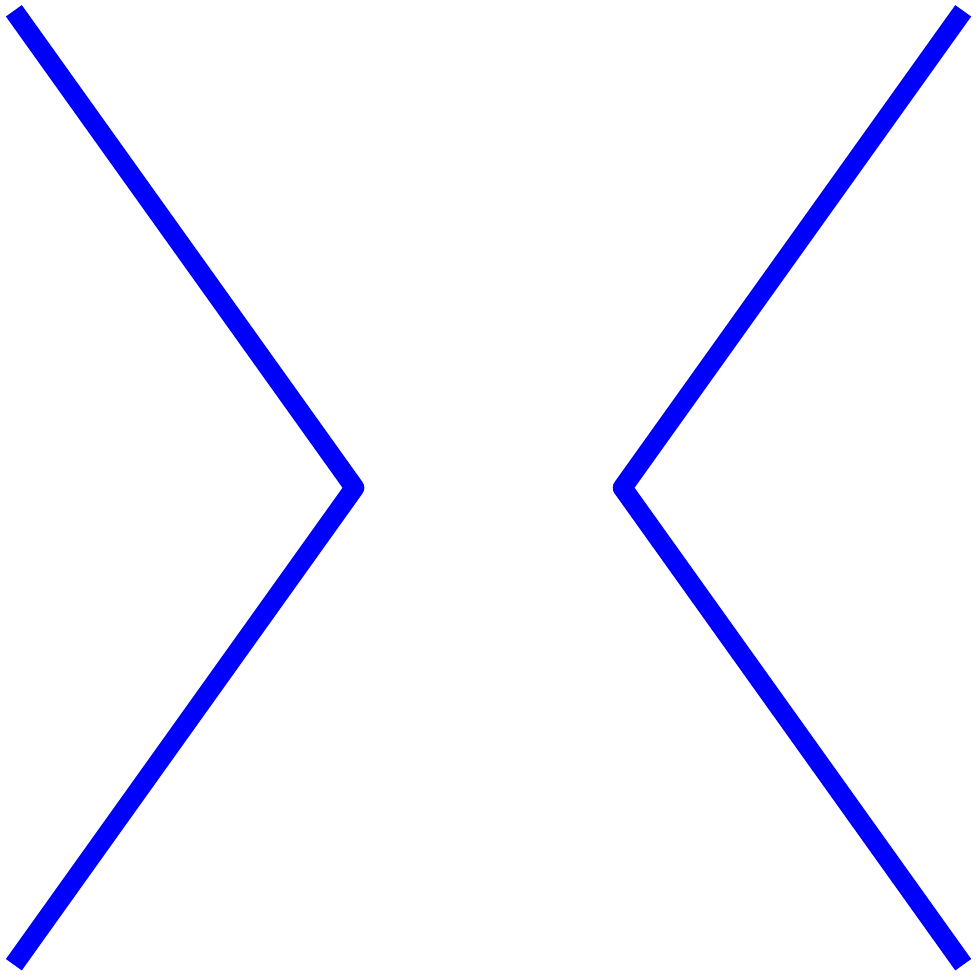} & \includegraphics[width=0.25in]{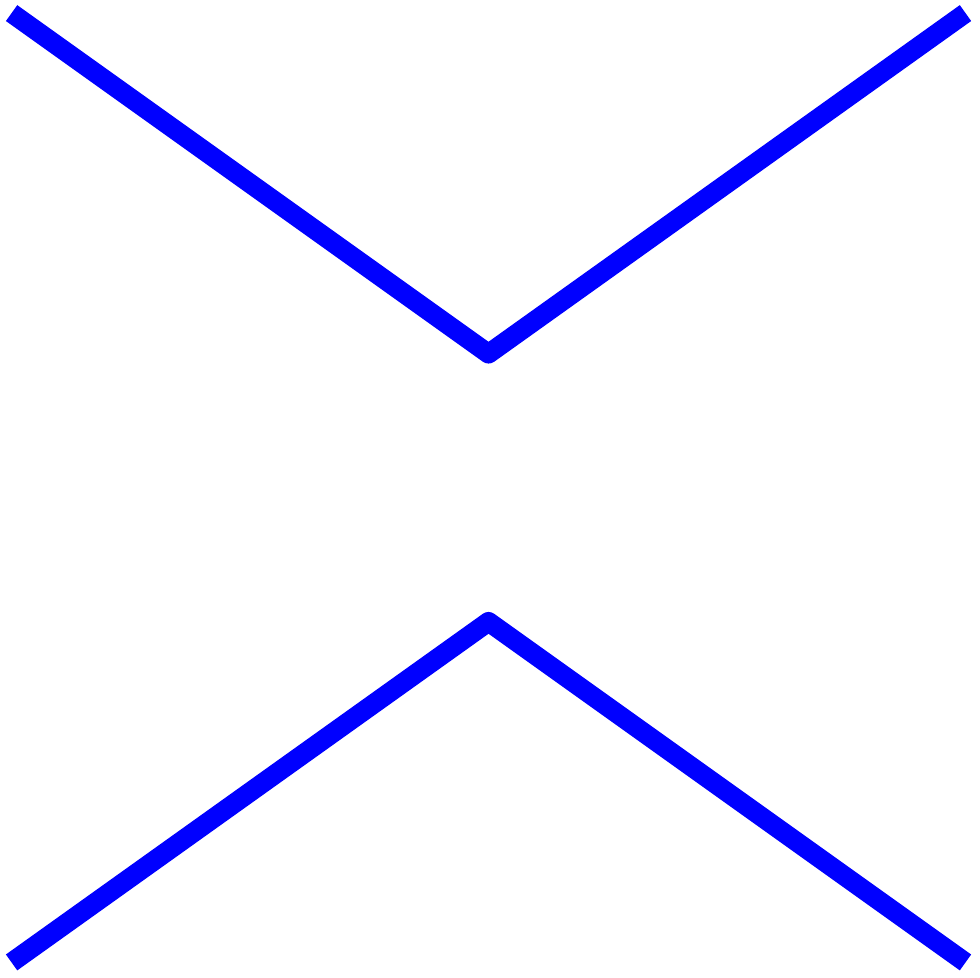} \\
         & & & &\\
         ZX Symbol  & \includegraphics[width=0.25in]{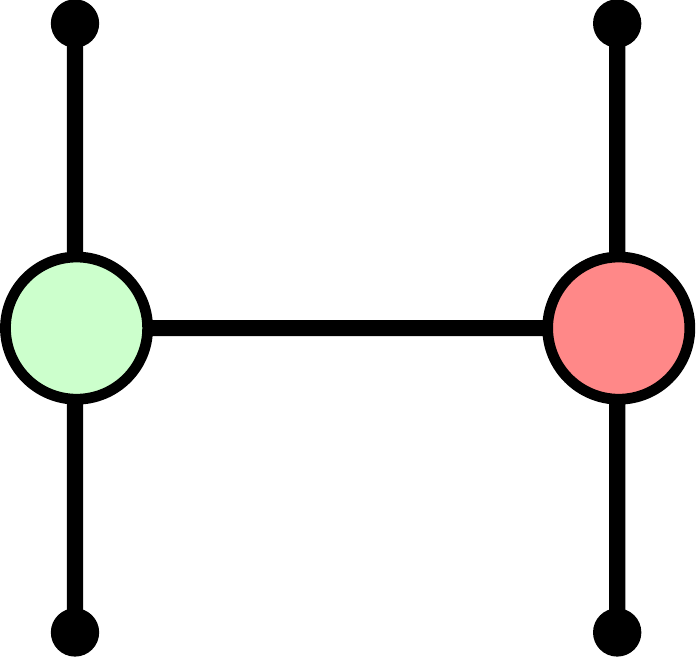} & \includegraphics[width=0.25in]{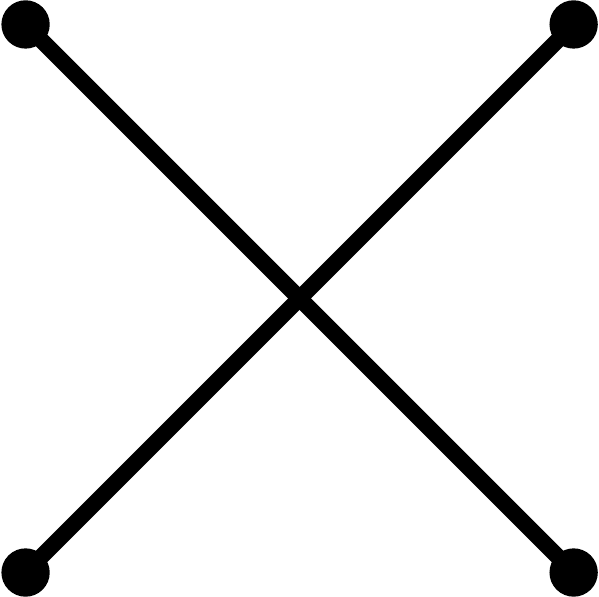} & \includegraphics[width=0.25in]{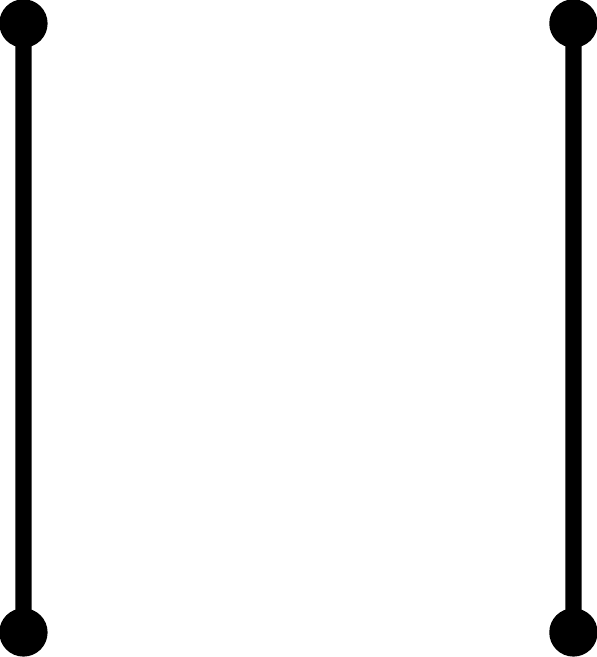} & \includegraphics[width=0.25in]{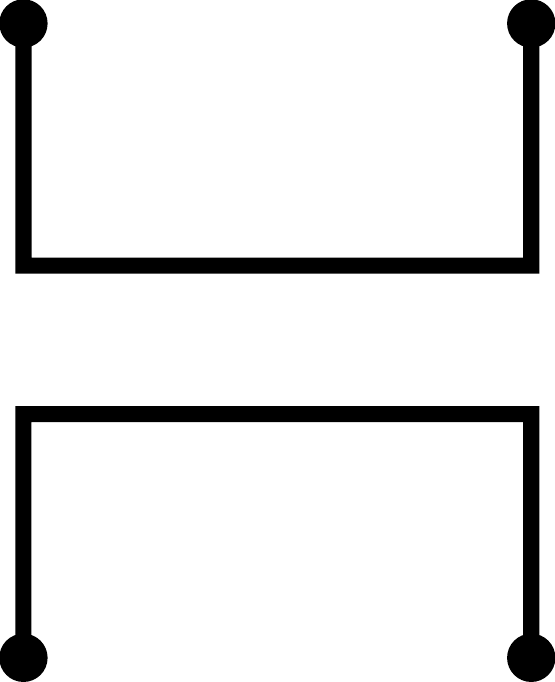} 
    \end{tabular}
    \caption{Summary of the probabilities (top row) associated with the different two qubit operations/measurements (middle row), and their ZX calculus symbols (bottom row).  }
    \label{tab:probs}
\end{table}

\begin{figure}
\includegraphics[width=1.05\linewidth]{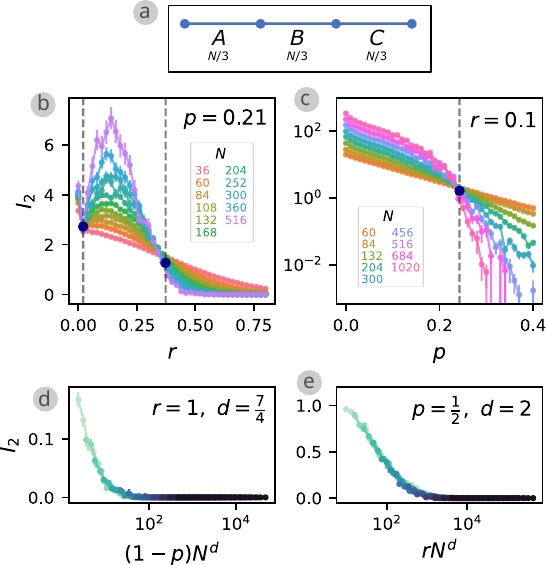}
    \caption{ (b) The bi-partite mutual information $I_2 = S_A + S_C - S_B$ at $p = 0.21$ as a function of $r$ for different system length between $N=36$ to $516$. The different curves for $I_2$ exhibit a crossing at the MPT, where they transition from linear growth to exponential decay with $N$. Two such crossing points at $r_{c1}$ and $r_{c2}$ are marked in (b). (c) $I_2$ at a fixed $r=0.1$ as a function of $p$ for system sizes between $N=60$ to $1020$, shows a transition point $p = 0.24$. (d)-(e) Finite size scaling of $I_2$ close to the phase diagram boundaries where the entanglement structure is governed by loop models in space-time; see Fig.~\ref{fig:phase_diag}.
    }
    \label{fig:MI}
\end{figure}

{\it Phase diagram.-} In order to map out the distinct phases associated with the dynamical circuit, we will use mutual information, $I_2 = S_A + S_C -S_B$, between two regions $A$ and $C$ separated by a third region $B$ (all of equal size) in a chain of total length $N$ with open boundary conditions [Fig.~\ref{fig:MI}(a)], where $S_A$ is the von Neumann entanglement entropy of subregion $A$. $I_2$ exhibits a crossing at the MPT, similar to the tripartite mutual information considered in Ref.~\cite{zabalo2020critical}. Namely, $I_2$ shows the following characteristic features as a function of $N$,
\begin{align}
 \label{mutualinfo}
\lim_{N\to \infty} I_{2} =\begin{cases}
    \exp(-a N) & \text{area law} \\
    \rm{constant} & \text{criticality} \\ 
    b N & \text{volume law}
\end{cases}\,.
\end{align}
In Fig.~\ref{fig:MI}(b)  we plot $I_2$ as a function of $r$ at a fixed $p = 0.21$ for different system sizes. 
There are two scale-invariant crossing points for $r$ representing two distinct critical points, $r_{c1}$ and $r_{c2}$. $I_2$ exhibits scaling collapse near both $r_{c1}$ and $r_{c2}$ with a critical exponent, $\nu = 4/3$ consistent with previous studies~\cite{li2018quantum,li2019measurement,zabalo2020critical,zabalo2022operator,gullans2020dynamical,li2021conformal}.
In panel (c), we also plot $I_2$ vs. $p$ for fixed $r=0.1$ showing a crossing at $p_c = 0.24$.

\begin{figure}
\includegraphics[width=1.0\linewidth]{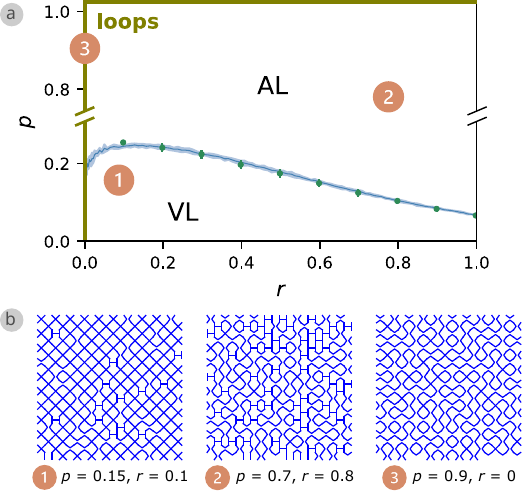}
     \caption{ (a) Phase diagram of the brickwork circuit with gates from Table~\ref{tab:probs}, controlled by probabilities $p$ and $r$. The area-law (AL) and volume-law (VL) phases are indicated. The solid blue line marks the MPT from mutual information (shading indicates error); see Eq.~\eqref{mutualinfo}. Green dots denote a classical percolation transition from ZX-calculus-based simplifications of Clifford circuits. khaki boundaries at $r=0$ and $p=1$ correspond to fully packed loop coverings of a square lattice. The coincidence of the MPT and  classical percolation transition suggest the universal behavior of the former is controlled by a hidden classical percolation transition.
(b) Three sample Clifford circuit realizations at the marked points on the phase diagram.}
     \label{fig:phase_diag}
\end{figure}

 There are asymptotic limits where the problem can be described quite simply, as shown in Fig.~\ref{fig:MI}(d)-(e), by perturbing away from the limit where CNOT gates are absent ($r=0$), or ($p=1$). In the latter case, the entanglement structure maps to classical lattice loops, as studied in monitored Majorana fermion models~\cite{nahum2020entanglement,merritt2023entanglement,KlockeMajoranaLoops}. Consider for example the point marked \textcircled{$3$} in Fig.~\ref{fig:phase_diag}(a) and its corresponding circuit realization in Fig.~\ref{fig:phase_diag}(b). 
In this case, the entanglement entropy across a bipartition corresponds to the number of open loops with both legs  emanating from the final time boundary and encircling the bipartition point, scaling as $\log N$ for loops without crossings ($p=1$) and $(\log N)^2$ for loops with crossings ($r=0$ and $p\ne 1$)~\cite{nahum2015deconfined}. The length of such loops in the bulk, $\ell$, scales as a fractal   $\ell\sim N^d$, where  $d=7/4$ and $d=2$ for non-crossing  and crossing loops, respectively.
We can verify these exponents by adding a small density of CNOT gates,  $\delta$, which have the leading effect of disentangling loops. Consequently, the loops are expected to disentangle when $\ell\sim 1/\delta $, which leads to a correlation length  $\xi \sim \delta^{-1/d}$, where $\delta=r(1-p)$ or $\delta=r$ in the vicinity  of the top and left boundaries of the phase diagram, respectively \cite{supp}. Indeed, $I_2$ shows universal scaling with the relevant exponents [see Fig.~\ref{fig:MI}(d)-(e)].
The boundary loop structures suggest that no classical percolation transition occurs anywhere in the phase diagram since the ``minimal-cut picture'' captures the logarithmic/squared logarithmic growth on the boundaries and the addition of CNOTs only increases the connectivity of the circuit structure within this picture.

In Fig.~\ref{fig:phase_diag}(a) we construct the phase diagram in the $r-p$ plane by extending our analysis of mutual information in Fig.~\ref{fig:MI} as a function of $p$ and $r$. The transition between the entangling volume law (VL) and disentangling area law (AL) regimes is marked by the blue line (shading represents error). 
We find that the phase boundary extends from the bottom $p=0$ boundary, showing a peak around $r=0.1$ and $p=0.24$. Interestingly, for smaller $r$ the phase boundary drops rapidly towards the origin at $p=0$ and $r=0$. We can understand this drop by noting that the critical loop phase and the entangling phase (VL) are necessarily separated by a disentangling phase (AL). This is a consequence of the disentangling effect of the CNOT gates acting on loops. Using a toy model of a random walker, we estimate the phase boundary asymptotically scales as $r_c(p) \sim \exp(-1/Ap)$ in the limit  $r\ll p\ll1$~\cite{supp}, where $A$ is constant. 
%\DC{Following statement necessary?} 
Thus, we conclude that our model exhibits a Clifford type MPT in the bulk of the phase diagram, which is not captured by the  minimal cut picture.

\begin{figure}
\includegraphics[width=1\linewidth]{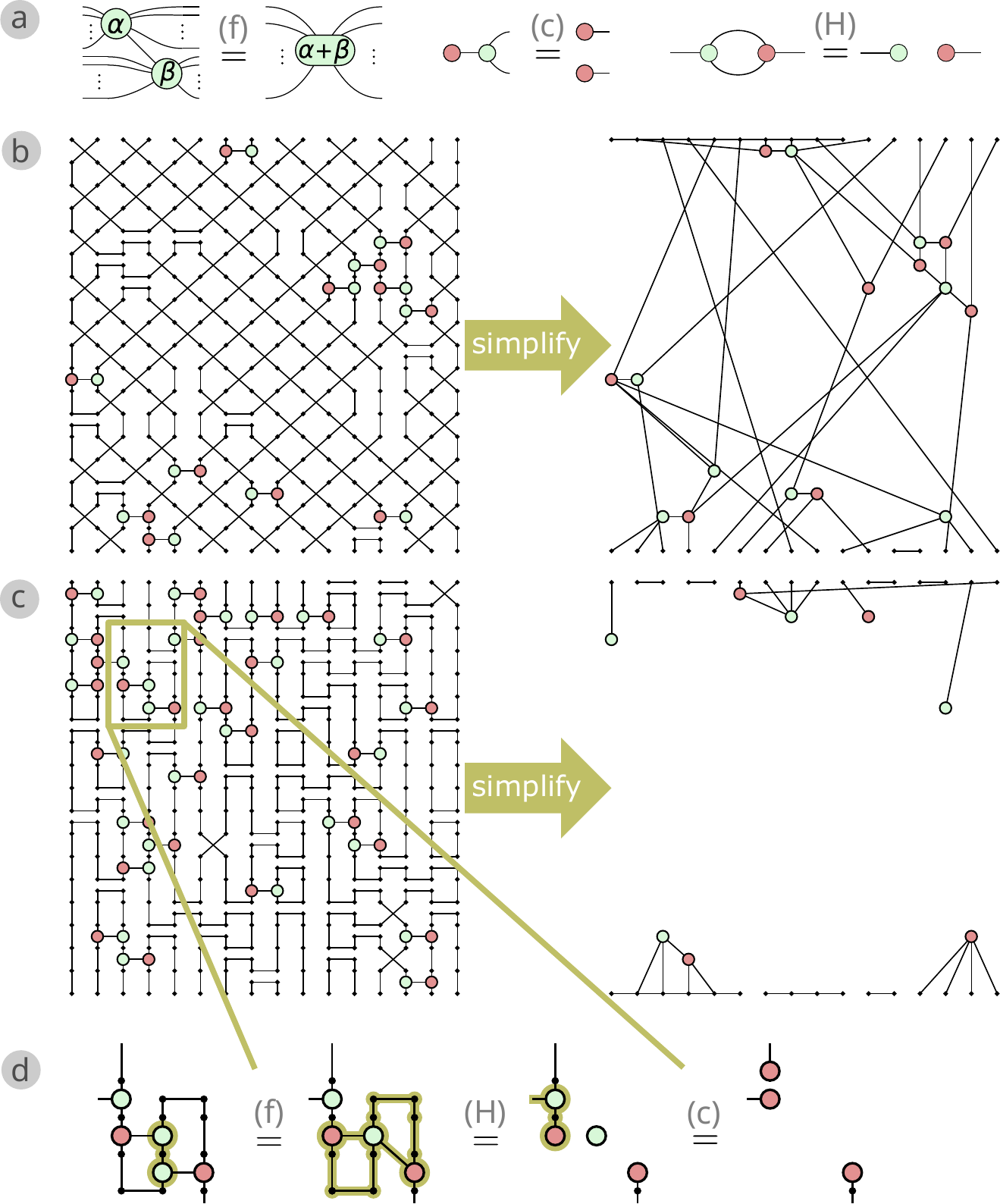}
     \caption{ 
     (a) Local ZX calculus simplification rules: (\emph{f}), (\emph{C}), and (\emph{H}) denote fusion, copy, and Hopf rules, respectively. (b)-(c) Left: ZX diagrams for circuits generated from Table~\ref{tab:probs} with $p=0.15$ \& $r=0.1$, and $p=0.7$ \& $r=0.8$, respectively. Right: Associated tensor networks resulting from the ZX simplification. (d) A specific region of circuit (c) simplified via successive implementation of the local rules.}
     \label{fig:ZX-rules}
\end{figure}

{\it ZX Calculus Enabled Circuit Simplification.-} ZX calculus is a universal framework for graphical representation of quantum processes~\cite{coecke2018picturing, Coecke_2011,Duncan2020graphtheoretic,van2020zx}, including quantum circuits and tensor networks. The basic building blocks consist of green ($Z$) Spiders \zx{\zxZ{\alpha}}, red ($X$) Spiders \zx{\zxX{\alpha}} and  wires that connect them.~\footnote{ZX diagrams generally also contain Hadamard boxes, which do not appear for our circuits.}
 The spiders correspond to the following Dirac notation of tensors
\begin{subequations}
\beq
\begin{ZX}
\leftManyDots{} \zxZ{\alpha} \rightManyDots{}
\end{ZX} 
& \equiv {\frac{1}{\sqrt{2}}}\left( \ket{0...0}\bra{0...0} + e^{i\alpha} \ket{1...1}\bra{1...1} \right)\,, \\
\begin{ZX}
\leftManyDots{} \zxX{\alpha} \rightManyDots{}
\end{ZX} 
& \equiv {\frac{1}{\sqrt{2}}}\left( \ket{+...+}\bra{+...+} + e^{i\alpha} \ket{-...-}\bra{-...-} \right)\,.
\eeq    
\end{subequations}
The number of wires on the right and left sides of the spiders corresponds to the dimensions of the \emph{bra}s and \emph{ket}s, which can be arbitrary. However, ZX-diagrams inherently lack a notion of ``flow direction,'' making this distinction artificial. 
The Greek letter in the spider stands for the phase angle; if absent, the angle is zero.

The following rules play a central role in our analysis: 
\emph{Spider fusion--} Two spiders of the same color can be fused and their angles are additive, corresponding to diagram (\emph f) in Fig.~\ref{fig:ZX-rules}(a).
\emph{Copy rule--} A single-legged spider of one color connected to a different colored three-legged spider is equivalent to two disconnected spiders with the color of the former, corresponding to diagram (\emph C) in Fig.~\ref{fig:ZX-rules}(a).
\emph{Hopf rule--} When different colored spiders are mutually connected by two wires, they decouple as shown in diagram (\emph H) of Fig.~\ref{fig:ZX-rules}(a). A successive application of the above rules to a circuit diagram can lead to a significant reduction in the number of spiders, and even disconnect nominally connected networks. 

We now explore the properties of ZX-simplified circuits initiated from the dynamical rules of  Table~\ref{tab:probs}. 
The only element in our model that has a non-trivial spider structure is the CNOT gate
\begin{align}
\text{CNOT} = \begin{ZX}[ampersand replacement=\&]
\zxN{} \rar \&[\zxwCol] \zxZ{} \ar[d] \rar \&[\zxwCol] \zxN{}\\
\zxN{} \rar \&[\zxwCol] \zxX{} \rar \&[\zxwCol] \zxN{}
\end{ZX}.
\end{align} 
In Fig.~\ref{fig:ZX-rules}(b)-(c), we show on the left the ZX representations for circuits generated for $p=0.15$, $r=0.1$ (i.e. in the entangling phase) and $p=0.7$, $r=0.8$ (i.e. in the disentangling phase), respectively.
With the upper and lower boundaries held fixed, we simplify the  diagrams \cite{supp} leading to the tensor network on the right.  
Notably, the initial and final times become decoupled in the disentangling phase, signaling a breakdown in information transmission through time, in contrast to the entangling phase, which preserves this information for an exponentially long duration.
For further insight, we have provided the simplification process tied to the highlighted region in Fig.~\ref{fig:ZX-rules}(c) in panel (d). The fusion rule (\emph{f}) first merges the two green spiders, followed by the disentangling using the Hopf rule (\emph{H}) at two separate locations. Finally, the copy rule (\emph{C}) applied in the upper left corner further disconnects wires.

\begin{figure}
\includegraphics[width=0.9\linewidth]{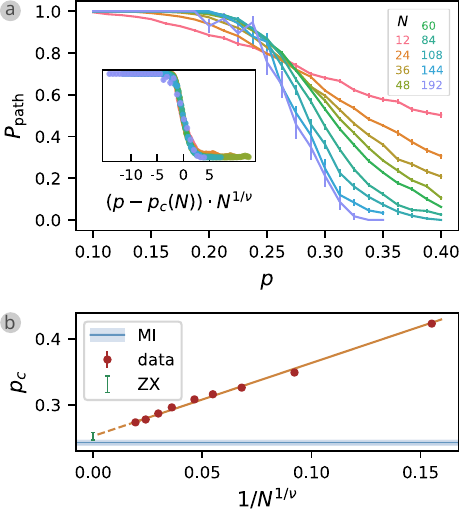}
\caption{(a) The probability ($P_{\text{path}}$) of finding a path between fixed initial and final times in the ZX simplified circuits for $r=0.1$ as a function of $p$ and $N$. Inset: Rescaled data, $P_{\text{path}} = f([p - p_c(N)]N^{1/\nu})$. (b) Critical $p_c(N)$ vs. $1/N^{1/\nu}$. The extrapolation $p_c(N\rightarrow\infty)$ coincides with the expected $p_c$ based on $I_2$ in Fig.~\ref{fig:phase_diag}. }\label{fig:findPathProb}
\end{figure}

We have computed the probability of finding a path connecting the initial and final times, $P_{\text{path}}$, by averaging over $10^6$ realizations of the ZX-simplified networks. 
We invoke the equivalence between spatial entanglement measures~\cite{skinner2019measurement,li2018quantum,chan2019unitary} and temporal measures of entanglement, such as  operator entanglement entropy~\cite{jonay2018coarsegraineddynamicsoperatorstate,nahum2021measurement} and purification~\cite{gullans2020dynamical} at the MPT. It is also worth noting that the dynamical critical exponent is $z=1$ in the $(1+1)-$dimensional models we are studying here~\cite{skinner2019measurement,li2018quantum,li2019measurement,li2021conformal,li2021statistical,zabalo2020critical,zabalo2022operator,gullans2020dynamical,kumar2024boundary}, implying a single critical exponent $\nu$ controls spatiotemporal correlations. Fig.~\ref{fig:findPathProb}(a) shows $P_{\text{path}}$ as a function of $p$ for $r = 0.1$ across various system sizes, and goes from unity to zero across a critical value, $p_c$, with a characteristic width, $\Delta p\sim 1/N^{1/\nu}$, where $\nu \approx 4/3$. Notably, $p_c$ decreases with increasing $N$ due to finite-size effects. This decrease reflects the enhanced success of the simplification algorithm in disconnecting circuits with growing system size in the disentangled phase.
The $N$-dependence of the threshold is also observed in the mean size of the  \emph{second-largest cluster} in the network \cite{supp}.  
%Overall, this trend suggests a shift in the critical value, highlighting the need for extrapolation to the thermodynamic limit to determine the critical threshold.

To determine the transition point, we employ a scaling ansatz, 
$P_{\text{path}}[p,N] = f\left(\left[p-p_c(N)\right]N^{1/\nu}\right)$, where $f(x)$ is a scaling function that satisfies $f(-\infty) = 1$ and $f(\infty) = 0$.
The rescaled data is shown in Fig.~\ref{fig:findPathProb}(a)-inset. In Fig.~\ref{fig:findPathProb}(b) we plot the resulting $p_c(N)$ vs. $N^{-1/\nu}$, and obtain $p_c(N\rightarrow\infty) = 0.251 \pm 0.009$. This value coincides with the critical threshold obtained from $I_2$ in Fig.~\ref{fig:MI}(b), where we found $p_c = 0.243\pm 0.003$.

The extrapolated $p_c(N)$ across the phase diagram is shown in Fig.~\ref{fig:phase_diag}(a) as green dots with error bars, indicating that the classical percolation transition coincides with the MPT. The structure of the simplified ZX diagrams allows us to compute the Hartley entropy via the minimal cut, which, for Clifford circuits, equals all R\'enyi entropies. This provides strong evidence that the MPT in Clifford circuits corresponds to a classical percolation transition, uncovered through our ZX-simplification algorithm.

However, an important clarification is in order here. Previously, Refs.~\cite{zabalo2020critical,zabalo2022operator,kumar2024boundary,li2021statistical} have reported deviations from classical percolation on the square lattice, seemingly at odds with our findings. However, the ZX-simplification is non-local, and the resulting network statistics differ from those of the original brickwork circuit with measurements. Hence, the critical properties need not be preserved through simplification~\cite{FisherMaNickel}. We have analyzed carefully the range of the simplification steps to support this renewed understanding of the transition \cite{supp}, and provide an open-source applet visualizing the procedure~\cite{Einatscode}.

\emph{Outlook--}
One of the central highlights of our study has been the demonstration of the power of ZX-calculus in simplifying complex quantum circuits into its essential skeletal structure, which has helped settle an important debate on the nature of the MPT in Clifford circuits. This opens up a number of interesting future directions. One of the technical challenges is to find efficient circuit simplification strategies in the case of generic random Clifford circuits, where further progress is likely possible by leveraging the power of machine-learning inspired algorithms \cite{estarellas,nagele2024optimizing}. An important future benchmark of our technique would be to demonstrate the critical properties of bulk~\cite{zabalo2022operator} and boundary~\cite{kumar2024boundary} operators based solely on the simplified networks. 
Another interesting direction is to study how symmetry, order and topology are reflected in the ZX-representation of circuits which generate non-trivial steady states~\cite{lavasani2021measurement,sang2021measurement,ippoliti2021entanglement}. This will help address the broader question of whether generic random Clifford circuits, with distinct symmetries, can be efficiently simplified using the setup discussed here. Finally, while we employed ZX-calculus to obtain a fundamental understanding of the Clifford-MPT, it may also provide a numerical advantage in two important future directions. The first is in the experimental realization of the MPT, where the simplification may help reduce circuit depth. The second, is perturbing away from Clifford circuits by adding a small density of T-gates, where ZX calculus can be extremely powerful~\cite{Duncan2020graphtheoretic}.

\emph{Acknowledgments.-} We thank Adam Nahum, Matan Ben Dov, Jed Pixley, Shlomo Havlin, and Anna Keselman for valuable discussions. We extend special thanks to Eran Sela and Daniel Azses for introducing us to ZX-calculus. JR and DC acknowledge funding from the US-Israel Binational Science Foundation under grant No. 2020213.

\emph{Data availability.-} The data that support the findings of this Letter are available in a public GitHub repository at Ref.~\cite{authors2025clifford}.

\bibliography{bibi}

\clearpage
\renewcommand{\thefigure}{S\arabic{figure}}
\renewcommand{\figurename}{Supplemental Figure}
\setcounter{figure}{0}
\begin{widetext}
\appendix
\begin{center}
{\bf \centering Supplementary Information for \\Unveiling a Hidden Percolation Transition in Monitored Clifford Circuits:\\ Inroads from ZX-Calculus}\\
{\centering Einat Buznach, Debanjan Chowdhury and Jonathan Ruhman }
\end{center}
%\maketitle

\section{The ``Minimal cut'' method to estimate the entanglement entropy across a bipartition}

The \emph{minimal cut} picture is a geometrical method to compute the Hartley function, $S_0$, in random unitary circuits~\cite{nahum2017quantum}. Thus, it provides an upper bound on the entanglement entropy. Let us consider a unitary circuit arranged in a brickwork manner, where we wish to estimate the entanglement between to subregions A and B, as shown in Fig.~\ref{fig:min_cut}. The argument relates $S_0$ to the \emph{minimal} number of ``cuts" required to partition the circuit into two separate parts. In the figure, the bonds corresponding to the cuts are marked in red, while the dashed line represents an arbitrary cut (which is not necessarily minimal). This cut starts at the bipartition point and can terminate at any of the boundaries or another bipartition point, with the red edges indicating the cut locations.

To understand why the number of cut bonds is related to the Hartley function we write the wave function as:
$$\ket{\psi}=\sum_{j=1}^{2^n}{\left(\sum_{\sigma_A}{A_{\sigma_A}^{j}\ket{\sigma_A}}\right)\otimes \left(\sum_{\sigma_B}{B_{\sigma_B}^{j}\ket{\sigma_B}}\right)}\,,$$
where $j=1,\ldots,2^n$ indexes the cut bonds, and $n$ is the number of cuts. This resembles the Schmidt decomposition as it is also a sum over product states belonging to the Hilbert spaces of the two parts. However, the rank of the Schmidt decomposition, $R=2^{S_0}$, is the \emph{minimal} number of product terms needed to reconstruct $\ket{\psi}$, implying $S_0 \leq n$. Ref.~\cite{nahum2017quantum} showed that minimizing over all cuts saturates this bound for Haar-random unitary gates, i.e., $\min_{\mathrm{cuts}} n = S_0$.

In the context of the MPT, it was shown in Ref.~\cite{skinner2019measurement} that single-site projective measurements effectively decimate bonds in the minimal cut picture. Let us denote the fraction of single site measurements in the circuit by $p$. When $p=0$, there are no measurements which means all edges exist, the network is well connected and every node is reachable from any other node, which means the system is in the volume law phase. As $p$ is increased, the bonds are randomly decimated. When $p$ is larger than a critical threshold $p_c$, the circuit falls into a large number of disconnected patches. In this case each qubit is only entangled with a small number of qubits in its close vicinity.  This is the area law phase. Thus, the minimal cut maps the transition in $S_0$ to a \emph{classical} bond percolation problem with a critical value $p = p_c$. 

However, we recall that $S_0$ is only an upper bound on the physical entanglement entropy $S_n$ (i.e. $S_m \leq S_n$ if $n\leq m$). 
Consequently, the MPT generally occurs at $p_q<p_c$ (see Fig. \ref{fig:phases_scheme}).
Indeed,  Refs.~\cite{skinner2019measurement,zabalo2020critical,nahum2021measurement,zabalo2022operator} found a regime $p_q<p<p_c$, where the circuit is still ``physically'' connected, but it does not generate long-ranged quantum correlations in space and time. 
In contrast, in the limit where the local Hilbert space dimension of the quantum degrees of freedom approaches infinity,\footnote{The degrees of freedom are generalized from qubits to qudits of dimension $q$ that is taken to infinity. In this case the fluctuation from one random matrix to another diminish and they tend to generate maximally entangled pairs with certainty~\cite{page1993average}.} the two transitions coincide, and the MPT becomes governed by classical percolation~\cite{jian2020measurement,bao2020theory}. One can interpret the percolation transition as a classical transition that precedes the quantum information transition as $p$ is decreased. This behavior is analogous to mean-field transitions, which typically precede the actual transition due to unaccounted-for fluctuations.

We also comment that it can be used to estimate other entanglement measures. For example, assume we wish to compute the operator entanglement entropy~\cite{jonay2018coarsegraineddynamicsoperatorstate}, which is defined as follows: Take the non-unitary evolution operator $\hat U$ and Schmidt decompose it into initial and final states 
\begin{align}
\hat U = \sum_{j=1}^{2^{S_0^{op}}}\mu_j  |j\rangle_f \langle j|_i
\end{align}
where $|j\rangle_{i,f}$ are the initial and final Schmidt states, respectively. Then we can define the operator entanglement entropy as
\begin{equation}\label{eq:app:S_op}
S^{op}_n = \frac{1}{1-n}\log_2\sum_{j=1}^{2^{S_0^{op}}} \tilde\mu_j^{2n} 
\end{equation}
where 
$$\tilde\mu_j = \frac{\mu_j}{\sqrt{\sum_{j=1}^{2^S_{0^{op}}}\mu_j^2}} $$
are the normalized singular values. The minimal cut that disconnects the initial and final times  in the circuit $\hat U$, as shown in Fig.~\ref{fig:min_cut}.b, is then the upper bound on $S_0^{op}$, which is saturated in the case of random Haar unitaries. At the MPT, both the bipartite and operator entanglement entropies undergo a transition~\cite{gullans2020dynamical,nahum2021measurement}. 
In the main text of this paper we invoke this minimal cut measure of the classical percolation transition in ZX-simplified of Clifford circuits, which precedes the MPT.

\begin{figure}
    \centering
    \begin{subfigure}[t]{0.45\textwidth}
        \centering
        \includegraphics[width=\textwidth]{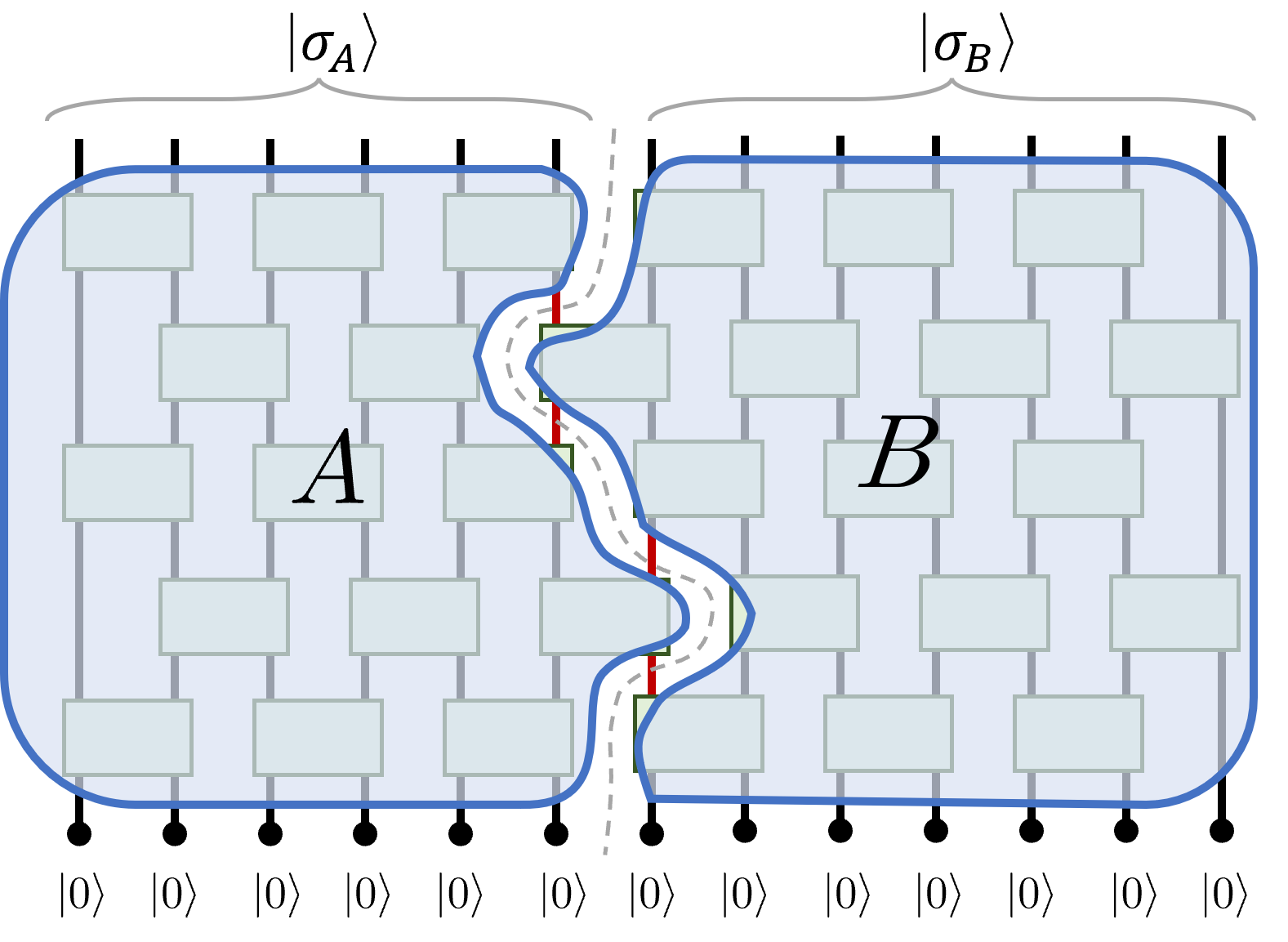}
        \caption{An illustration of an arbitrary cut. The cut is presented by the gray line, and the red lines are the bonds which were crossed by this cut. The tensors $A$ and $B$ are obtained by summing all the internal bonds and take the initial null state to the states $\ket{\sigma_A}$ and $\ket{\sigma_B}$, which are in the Hilbert space of subsystem A and B, respectively.}
    \end{subfigure}
    \hfill
    \begin{subfigure}[t]{0.45\textwidth}
        \centering
        \includegraphics[width=\textwidth]{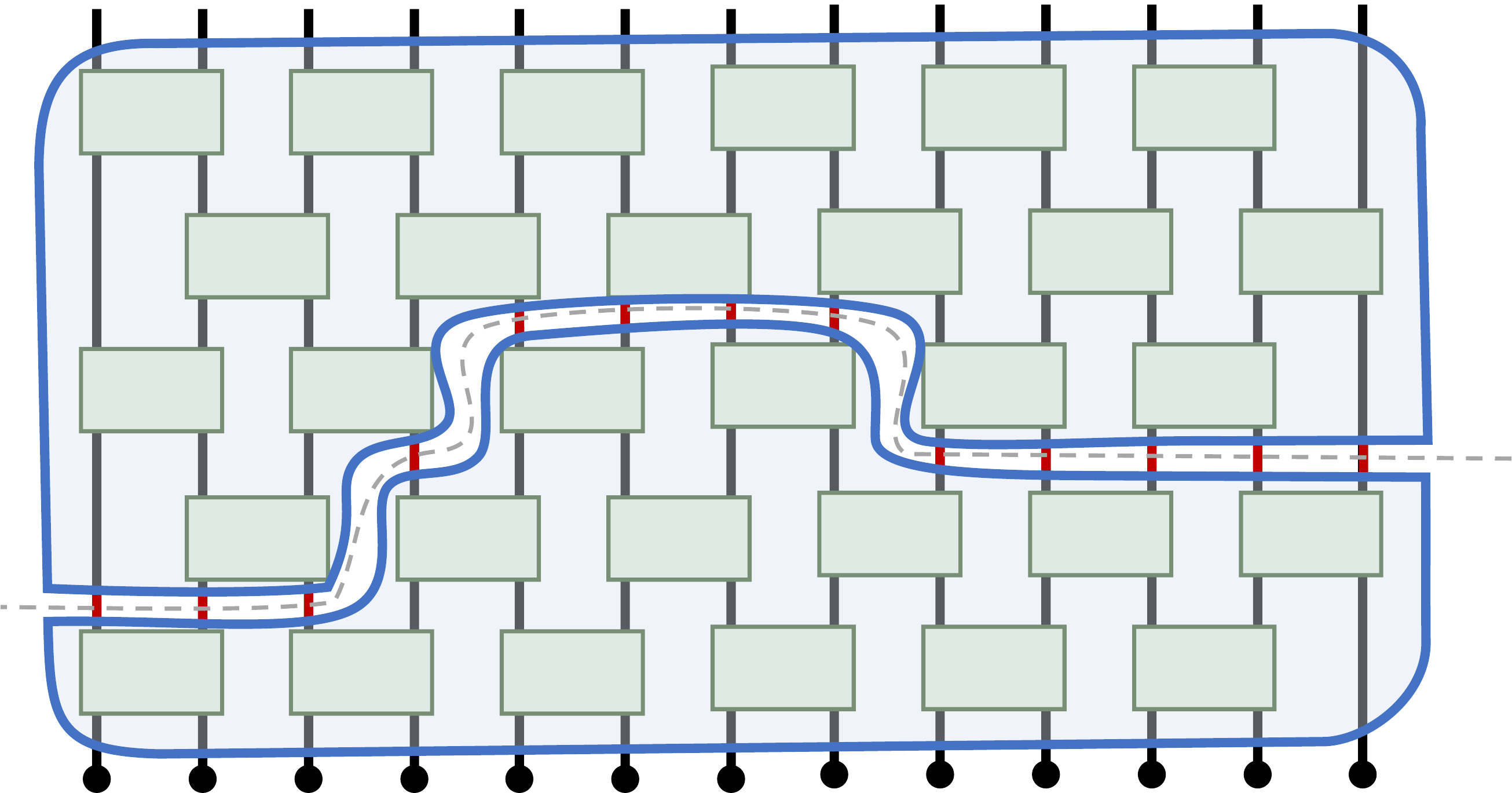}
        \caption{The $n=0$ limit of the operator entanglement entropy Eq.~\eqref{eq:app:S_op} is captured by a minimal cut in time rather than space. At the percolation threshold the free energy of both vertical and lateral cuts vanishes. Indeed, the dynamical critical exponent is $z=1$. In the main text we analyze the classical connectivity of Clifford circuits between initial and final times. Thus, we are alluding to the $n=0$ limit of the operator entanglement entropy. Given that the entanglement spectrum is flat  for Clifford circuits we anticipate this transition to be the MPT. }
    \end{subfigure}
    \caption{}
    \label{fig:min_cut}
\end{figure}

\begin{figure}
    \centering
    \includegraphics[width=0.6\linewidth]{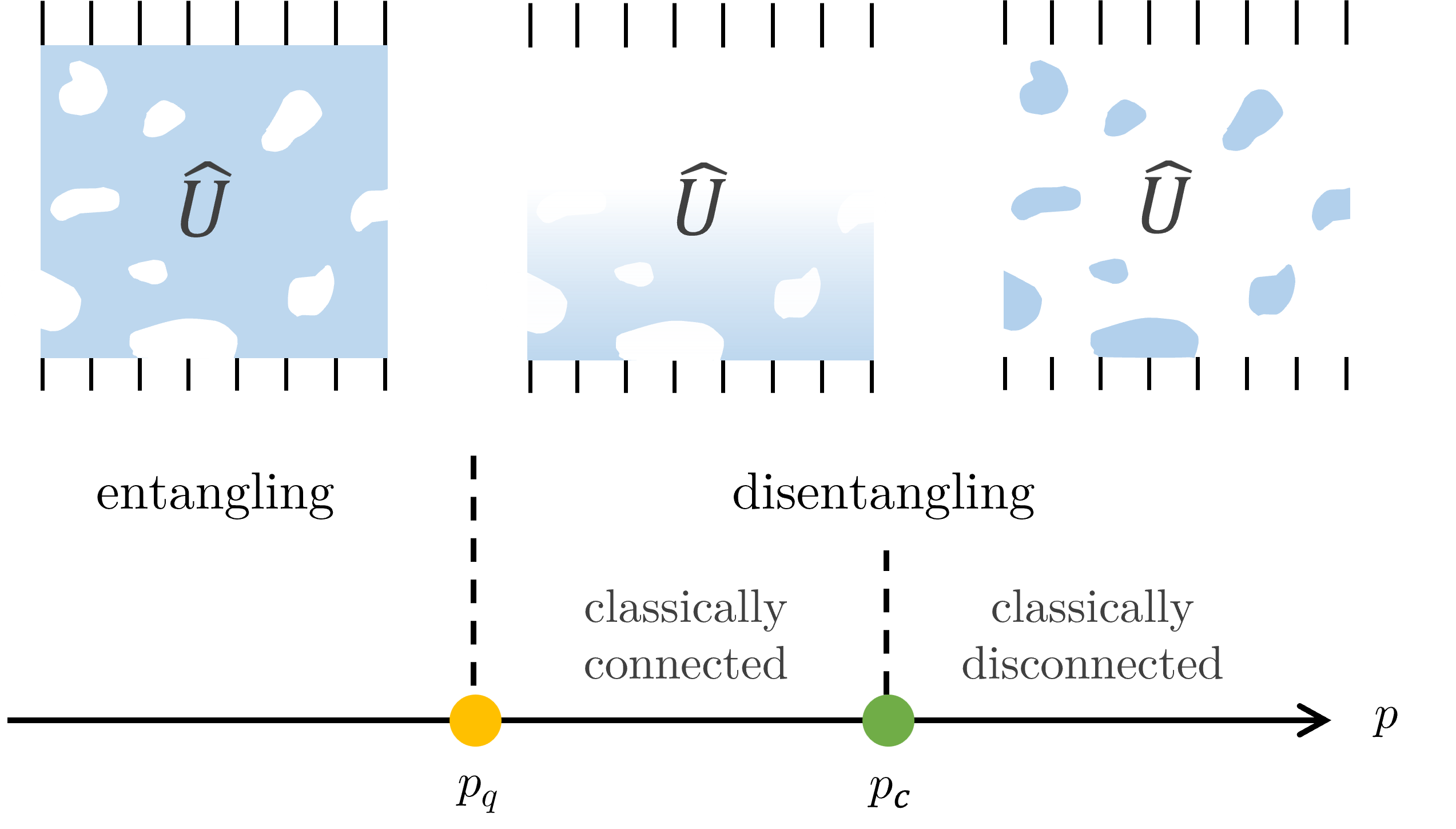}
    \caption{Typical phase diagram. A random unitary circuit is interspersed with measurements, which appear at a density $p$ in space-time. $p_c$ marks the classical percolation threshold of the circuit if measurements are associated with decimated bonds. At this threshold the Hartley function $S_0$ undergoes a phase transition from VL to AL. $p_q$ marks the MPT from the entangling phase (VL of the von Neumann entropy $S_1$) to the disentangling phase (AL), controlled by $S_1 < S_0$. 
    For finite Hilbert space, $p_q<p_c$, which results in 3 regions, as shown (On  both sides of the MPT $S_0$ is on its VL phase).
    An intuitive visualization is offered above the $p$ axis:  $\hat{U}$  represent the circuit, where  light blue regions are unmeasured and  nodes which have been measured are white. The circuit of $p_q<p<p_c$ is connected, but the can not transmit quantum information. Thus it is ``fading away.''
    }
    \label{fig:phases_scheme}
\end{figure}

\section{Initial State Dependence}

\begin{figure}
    \centering
    \begin{subfigure}[t]{0.65\textwidth}
        \centering
        \includegraphics[width=\textwidth]{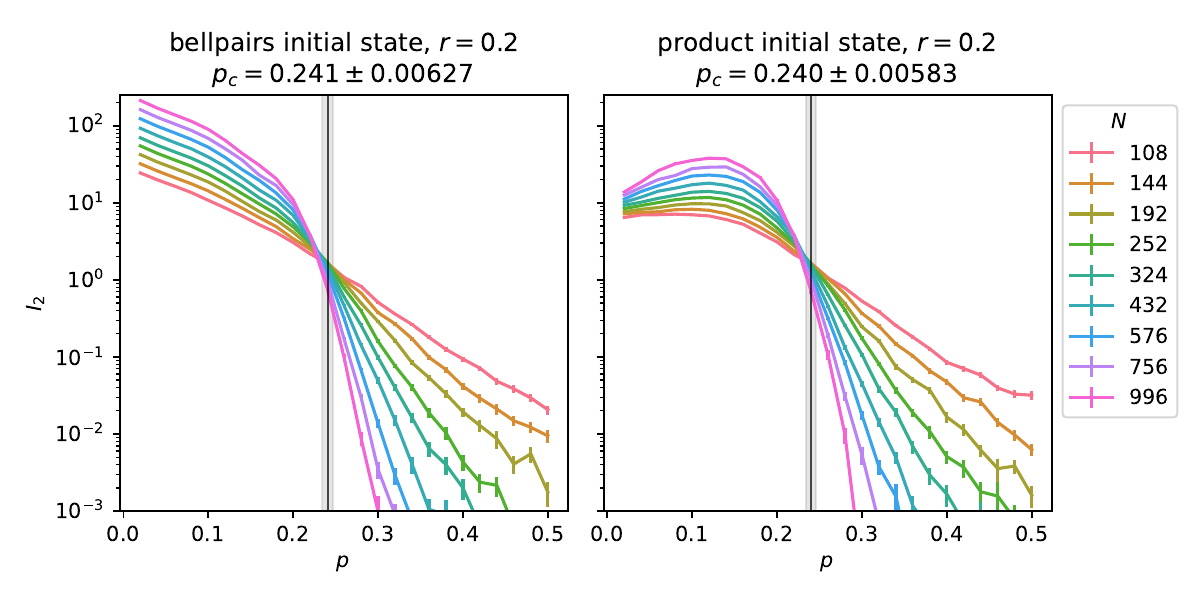}
        \caption{$I_2(p)$ for $r=0.2$ and varying system sizes $N$. The horizontal line marks the critical value $p_c$. While the behvior of the mutual information in the volume-law is different, both initial state preparations yield the same critical transition.}
    \end{subfigure}
    \hfill
    \begin{subfigure}[t]{0.3\textwidth}
        \centering
        \includegraphics[width=\textwidth]{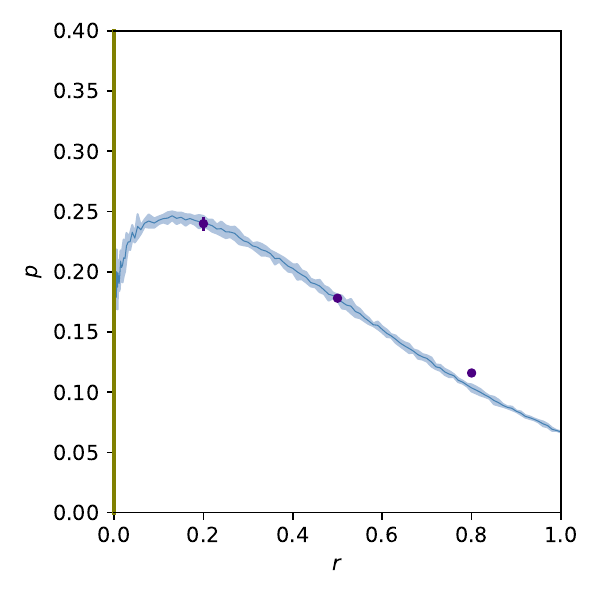}
        \caption{Phase diagram from the main text with three representative points confirming that phase boundaries remain consistent under product state initialization.}
    \end{subfigure}
    \caption{Effects of initial state choice on phase transitions in monitored Clifford circuits.}
    \label{fig:initial_states}
\end{figure}

Our model initializes the system with a product of neighboring Bell pairs, as described in the main text. Here, we show that this choice does not significantly alter the phase diagram. Figure~\ref{fig:initial_states}(a) compares the mutual information as a function of $p$ for $r=0.2$ across different system sizes $N$ at $t=4N$, revealing that the critical point $p_c$ remains unchanged. This robustness is further evidenced in Figure~\ref{fig:initial_states}(b), where three representative points in the $(p,r)$ parameter space confirm that the phase boundaries are largely preserved under product state initialization.

\section{Additional results for mutual information}
In the main text we have discussed the mutual information 
\begin{align}I_{2} = S_{A}+S_C - S_B\end{align} 
between two equal sized regions of size $N/3$, $A$ and $C$ , separated by a third equal sized region $B$, in a system with open boundary conditions, as shown in Fig.~\ref{fig:MI}.(a).

The mutual information is advantageous in determining the location of the MPT because it is scale invariant at the critical point and exhibits the following $N$-dependence
\begin{align}
 \label{eq:SI:mutualinfo}
\lim_{N\to \infty} I_{2} =\begin{cases}
    \exp(-a N) & \text{area law} \\
    \rm{constant} & \text{criticality} \\ 
    b N & \text{volume law}
\end{cases}\,.
\end{align}
The minimal cut picture  provides intuition regarding this dependence. Namely, in the volume law phase and $t\to\infty$ the minimal cuts of the regions $A$ and $C$ run side ways, while the minimal cut of region $B$ runs between the two boundaries in the bulk, such that  $S_A = S_C =S_B = I_2= N/3$. In the area law phase the contribution of each minimal cut is only coming from the small vicinity near the biparition points such that $S_A + S_C = S_B$ and $I_2 = 0$. The transition between these two behaviors forces $I_2$ to cross at $p=p_c$. 

Numerically, the transition locating process is similar to Binders cumulant. We find the crossing points of pairs of successive $N$'s. They seem to have no dependence on $N$, so the transition point is set to be weighted average of these points.

In the main text Fig.~\ref{fig:MI} (b)-(c) we plot the mutual information across the transition. In Fig.~\ref{fig:pcuts} and Fig.~\ref{fig:rcuts} we show that the transition is consistent with the classical percolation exponent $\nu = 4/3$.

\begin{figure}
    \centering
    \includegraphics[width=\linewidth]{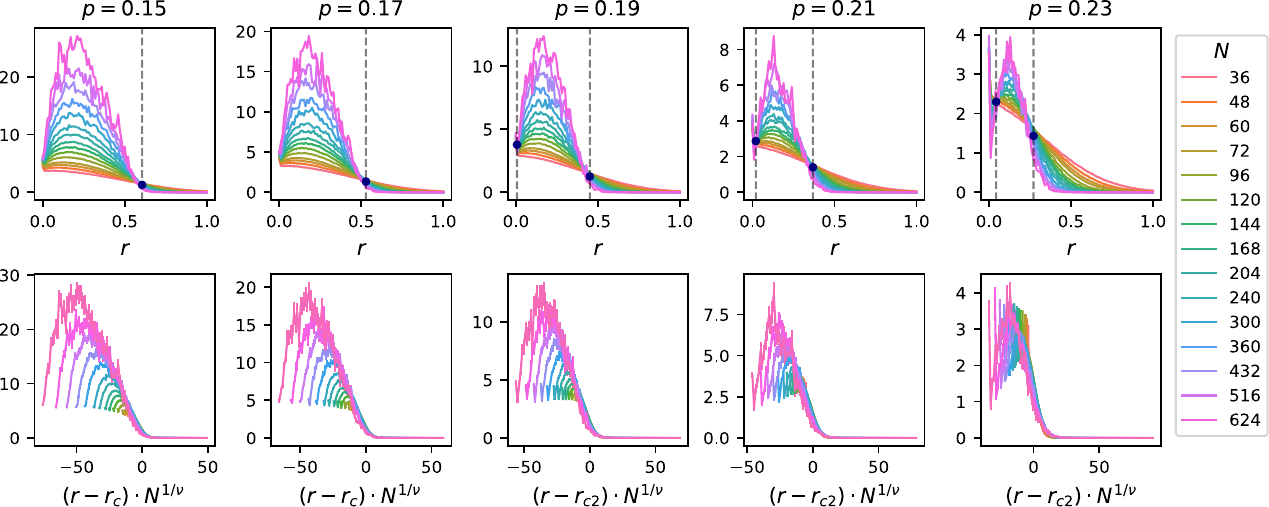}
    \caption{top: mutual information $I_2$ for constant $p$ values and changing $r$. bottom: scaling for $r_c$ identified in the matching top pane, and $\nu=\frac{4}{3}$}
    \label{fig:pcuts}
\end{figure}

\begin{figure}
    \centering
    \includegraphics[width=\linewidth]{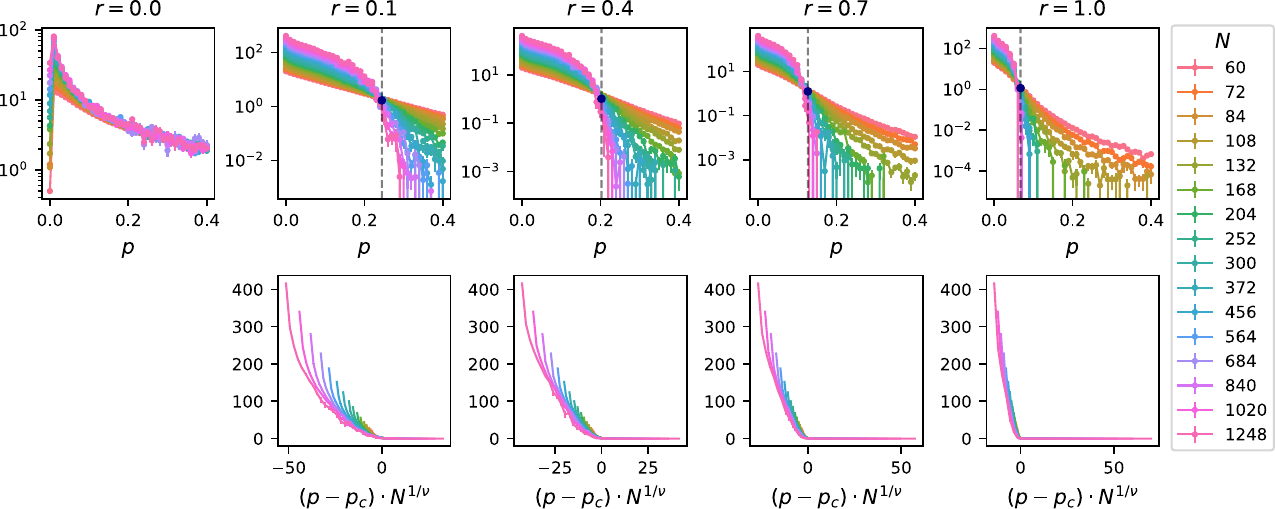}
    \caption{top: mutual information $I_2$ for constant $r$ values and changing $p$. bottom: scaling for $p_c$ identified in the matching top pane, and $\nu=\frac{4}{3}$}
    \label{fig:rcuts}
\end{figure}

\section{Heuristic arguments for the transition close to loop models}
In the main text we have discussed the limits of $p=1$ and $r=0$ (see Table~\ref{tab:probs}), which realize loop coverings on the square lattice, without and with crossings, respectively. In both cases the state, at any given time, is a product state of Bell pairs pairs of qubits, e.g. $i$ and $j$
\begin{align}\label{eq:supp:BP}
& |\Psi\rangle = {\tens_{s(i,j)}|BP\rangle_{ij}}
\\&|BP\rangle_{ij} \in  \begin{cases}
\frac{1}{\sqrt{2}}(|00\rangle_{ij} +|11\rangle_{ij})\\
\frac{1}{\sqrt{2}}(|01\rangle_{ij} +|10\rangle_{ij})\\
\frac{1}{\sqrt{2}}(|00\rangle_{ij} -|11\rangle_{ij})\\
\frac{1}{\sqrt{2}}(|01\rangle_{ij} -|10\rangle_{ij})
\end{cases}
\end{align}
where $s(i,j)$ runs over a specific random configuration of $N/2$ such pairs. 
The entanglement across a bipartition (the cut) is then given by the number of pairs, such that $i$ is on one side of the bipartition and $j$ is on the other. 

Using the vast knowledge about the statistical properties of such loops~\cite{cardy2005sle,nahum2015deconfined}, Nahum and Skinner~\cite{nahum2020entanglement} have predicted that the entanglement of a bipartition grows with system size like $\sim\log N$ and $\sim(\log N)^2$   for loops without and with crossings, respectively~\footnote{One can also cutoff the logarithmic growth using dimerization; modulating an imbalance between Bell pair measurement  and identity gate density on even and odd bonds, which drives the system into an area law phase~\cite{nahum2015deconfined,merritt2023entanglement}.}. Let us quickly follow their argumentation.  
The logarithmic growth is the result of the power law distribution of length of arcs~\cite{nahum2020entanglement}
\begin{align}
    P(L)\sim \begin{cases}
        \frac{1}{L^2} & \text{loops without crossings} \\
        \frac{\log L}{L^2} & \text{loops with crossings} 
    \end{cases}
\end{align}
which is the distance between the two endpoints of loops that terminate at the final time boundary. 
The entanglement of a sub region of size $N$ is related to the integral over the number of loops that emanate  inside the region and terminate outside of it
\begin{align}
S(N)\sim \sum_{x=1}^N  \sum_{L=N-x+1}^\infty  P(L)\,,
\end{align}
which gives the aforementioned asymptotic behavior. 

In the bulk, the length of  these loops, $\ell$, is distributed algebraically  
$P(\ell) \sim \ell^{-\tau}$, with $\tau = 1+2/d_f$ and $d_f$ is the fractal dimension of the loop. For loops without crossings $d_f = 7/4$ and for loops with crossings $d_f = 2$. In the latter case there are logarithmic corrections~\cite{nahum2015deconfined}. The above distribution corresponds to randomly drawing a loop out of the group of all loops with uniform probability. However, if we select a random bond on the lattice and consider the distribution of loops  emanating from it, there is a bias towards larger loops because they take a larger volume of sites/bonds. Consequently, the distribution of loop lengths emanating form a random site is given by $\sim \ell P(\ell)$, which implies the average loop length diverges.

To analyze the effect of CNOT gates on top of this loop structure we note that any CNOT gate connecting two different points of the same loop (i.e. linking the loop to itself) will dissect the loop and disentangle the Bell pair associated with its end points. This can be understood  by applying the CNOT gate to any of the four possible states in Eq.~\eqref{eq:supp:BP}, which results in a product state of the two qubits.  Conversely, a CNOT gate can also link two different loops,  leading to a four qubit entangled state and thus possibly enhance the entanglement across the cut. 
However, we may conjecture that near the critical phase the probability of a CNOT gate linking a loop to itself is greater than the probability of linking two different loops in a way that enhances the entanglement across the cut point. 
Linking a loop to itself at two different locations should scale as $\sim r \ell \sim r N^{1/d_f}$. This is because the number of points where a loop crosses itself (or nearly crosses itself) is propositional to its length. Anytime a  crossing or a near crossing point is exchanged by a CNOT gate the loop is disentangled and no longer contributes to the entanglement entropy.

\begin{figure}[htbp]
    \centering

    % Left panel (previously wrapped figure)
    \begin{subfigure}[t]{0.39\textwidth}
        \centering
        \includegraphics[width=\linewidth]{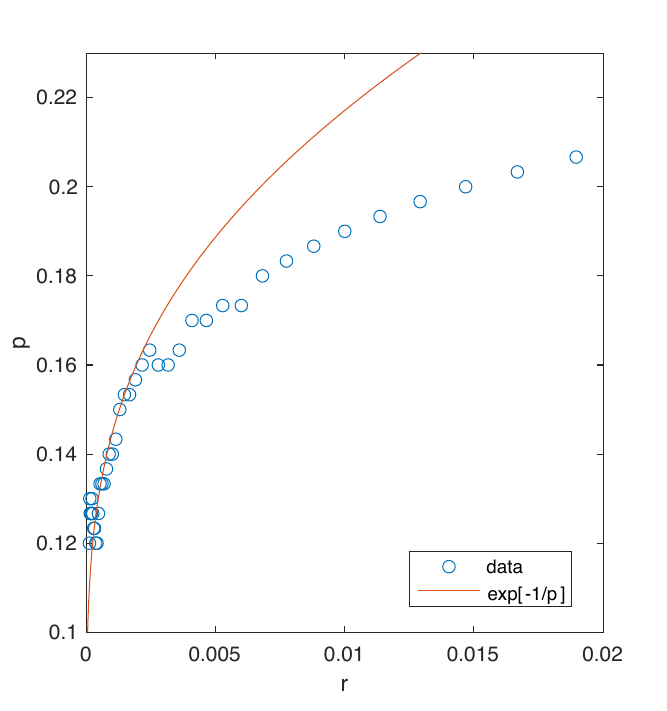}
        \caption{Comparison between the phase boundary obtained from the mutual information crossing point and the exponential curve obtained from the heuristic picture of percolating loops, Eq.~\eqref{eq:sup:r(p)}.}
        \label{fig:sup:small_p_r_fit}
    \end{subfigure}
    \hfill
    % Right panel (previous figure)
    \begin{subfigure}[t]{0.56\textwidth}
        \centering
        \includegraphics[width=\linewidth]{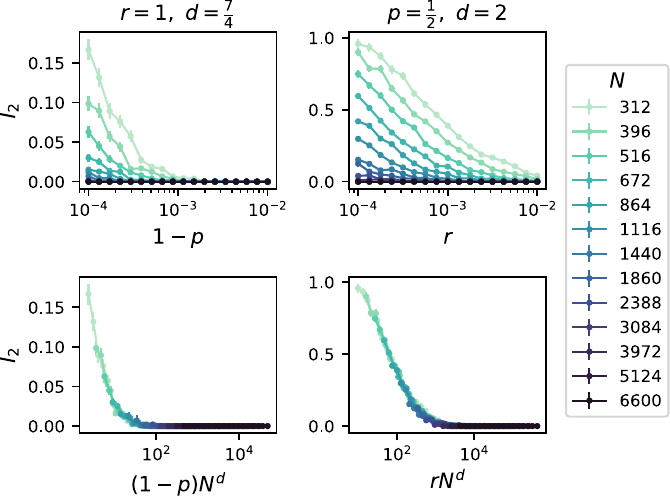}
        \caption{The top two panels show the raw data at $r=1$ and as a function of $1-p$, and at $p=1/2$ and as a function of $r$. The two bottom panels show the same data, where the system sizes are rescaled with the dimension $d=7/4$ for the left panel and $d=2$ for the right panel.}
        \label{fig:loops}
    \end{subfigure}

    \caption{(Left) Comparison of the phase boundary to the heuristic exponential fit. (Right) Loop data with and without scaling collapse.}
\end{figure}

We can verify the above conjecture  by adding a small density of CNOT gates in the vicinity of the critical loop phases on the boundaries $r = 0$ and $p=1$ in our simulations. The prediction is that loops of length $\ell$ grater than $1/r$ will be disentangled with probability one. Thus, as the system size is increased the logarithmic/logarithmic-squared growth will be cutoff by the scale 
$$\xi \sim r^{-1/d}\,.$$ 
We qualitatively observe this behavior directly in the bipartite entanglement entropy, which grows logarithmically with system size until it saturates above $\xi$. However, the saturation regime is wide, which hampers an accurate  extraction of $\xi$. In Fig.~\ref{fig:loops} we plot the mutual information between two equal sized regions separated by a third equal sized region in a system with open boundary conditions. 
The top two panels show the raw data at $r=1$ and as a function of $1-p$, and at $p=1/2$ and as a function of $r$. The two bottom panels show the same data, where the system sizes are rescaled with the dimension $d$. 
These curves exhibit  universal scaling with the above correlation length. This establishes the disentangling effect of the CNOT gates. It implies that the critical loop models are unstable to an infinitesimal density of CNOT gates. Consequently, the immediate vicinity of the boundaries $p=1$ and $r=0$ is an area law phase. 

With this heuristic picture in hand we turn to estimate the transition point between area law to volume law upon increasing $r$ further. We start with the limit of $r=0$ and $p\ll1$. As mentioned the average loop length diverges. Thus, to estimate a typical size of loops in the small $p$ limit we use a toy model of paths of a random walker, where the $p$ fraction of sites occupied by Bell pair measurements or identity matrices take the role of  scattering sites. Assuming each step is a unit of time the diffusion  constant is given by the distance between scatterers, $D = 1/p$. We conjecture that the typical loop size, $\bar \ell$, is obtained by integrating the probability of being at the origin until it reaches order 1
\begin{align}\label{eq:sup:P_r}
\int_{t_0}^{\bar \ell}dt \frac{A}{D t}  = {A p} \log \frac{\bar \ell}{t_0}\sim 1\,,
\end{align}
where $A$ is a constant and $t_0$ is the UV cutoff, naively set by the step size $t_0 \sim 1$. 
In this case, when $r \bar \ell \sim 1$ there is at least one CNOT gate per typical loop and we expect percolation to occur. This leads to the prediction 
\begin{align}\label{eq:sup:r(p)}
r\sim \exp(-1/Ap)\,.
\end{align}

In Fig.~\ref{fig:sup:small_p_r_fit} we compare Eq.~\eqref{eq:sup:r(p)} with the data in the limit of $r\ll p\ll 1$. The data is consistent with the above exponential behavior. However, the fit is only applicable in a finite region, and we cannot establish the asymptotic essential singularity. This is because it becomes increasingly challenging to obtain the transition point with smaller $r$ and $p$. Namely, the system size $N$ must be greater than $N \gg 1/r \sim \exp(1/Ap)$ to include more than one CNOT gate per realization. We may conclude that the data is consistent with Eq.~\eqref{eq:sup:r(p)}, but not more than that.

\section{Review of ZX-Calculus}
In this section, we review key elements of ZX-calculus~\cite{coecke2018picturing, Coecke_2011} presented in the main text. ZX-calculus is a powerful graphical framework for representing and manipulating quantum processes, including quantum circuits. The ZX-calculus uses diagrams composed of Z spiders \zx{\zxZ{}} and X spiders \zx{\zxX{}} connected by wires, potentially decorated with Hadamard gates \zx{\zxNone{} \rar & \zxH{} \rar & \zxNone{}}. The number of incoming/outgoing wires attached to a spider indicates the number of qubits it acts on. These diagrams can represent quantum states, operators, and circuits, For example:
\begin{align}
\label{eq:zx-examples}
 \begin{ZX}[ampersand replacement=\&]
\zxZ*{} \rar \&[\zxwCol] \zxN{}
\end{ZX}
 &= \ket{0}+\ket{1}\propto\ket{+}\\
 \begin{ZX}[ampersand replacement=\&]
\zxN{} \rar \&[\zxwCol] \zxZ*{\pi}
\end{ZX}
&= \bra{0}-\bra{1}\propto\bra{-}\\
\begin{ZX}[ampersand replacement=\&]
\zxN{} \rar \&[\zxwCol] \zxX*{\pi} \rar \&[\zxwCol] \zxN{}
\end{ZX}
&= \ket{+}\bra{+}+e^{i\pi}\ket{-}\bra{-}=X\,,
\end{align}
where an empty spider indicates a phase of $2\pi$.
The entire Clifford group is generated by CNOT, Hadamard, and S gates.  In ZX notation these assume the form
\begin{align}
\text{CNOT} = \begin{ZX}[ampersand replacement=\&]
\zxN{} \rar \&[\zxwCol] \zxZ{} \ar[d] \rar \&[\zxwCol] \zxN{}\\
\zxN{} \rar \&[\zxwCol] \zxX{} \rar \&[\zxwCol] \zxN{}
\end{ZX},\qquad
\text{S}= \begin{ZX}[ampersand replacement=\&]
\zxN{} \rar \&[\zxwCol] \zxZ*{\frac{\pi}{2}} \rar \&[\zxwCol] \zxN{}
\end{ZX},\qquad
\text{H}=\begin{ZX}[ampersand replacement=\&]
\zxN{} \rar \&[\zxwCol] \zxH{} \rar \&[\zxwCol] \zxN{}
\end{ZX}
\end{align}
Thus, for the circuits generated by the gates listed in Table~\ref{tab:probs}, only the CNOT gate is modified.
Note that when replacing S = \begin{ZX}[ampersand replacement=\&]
\zxN{} \rar \&[\zxwCol] \zxZ*{\frac{\pi}{2}} \rar \&[\zxwCol] \zxN{}
\end{ZX} with T = \begin{ZX}[ampersand replacement=\&]
\zxN{} \rar \&[\zxwCol] \zxZ*{\frac{\pi}{4}} \rar \&[\zxwCol] \zxN{}
\end{ZX}, a base for universal computation is achieved. T gates are typically more challenging to implement in a fault-tolerant fashion compared to Clifford gates. For this reason, the number of T gates (T-count), is often used in quantum computing for resource estimation, algorithm complexity comparison, and circuit optimization
~\cite{Kissinger2020red}.

\begin{figure}
    \centering
    \includegraphics[width=0.7\linewidth]{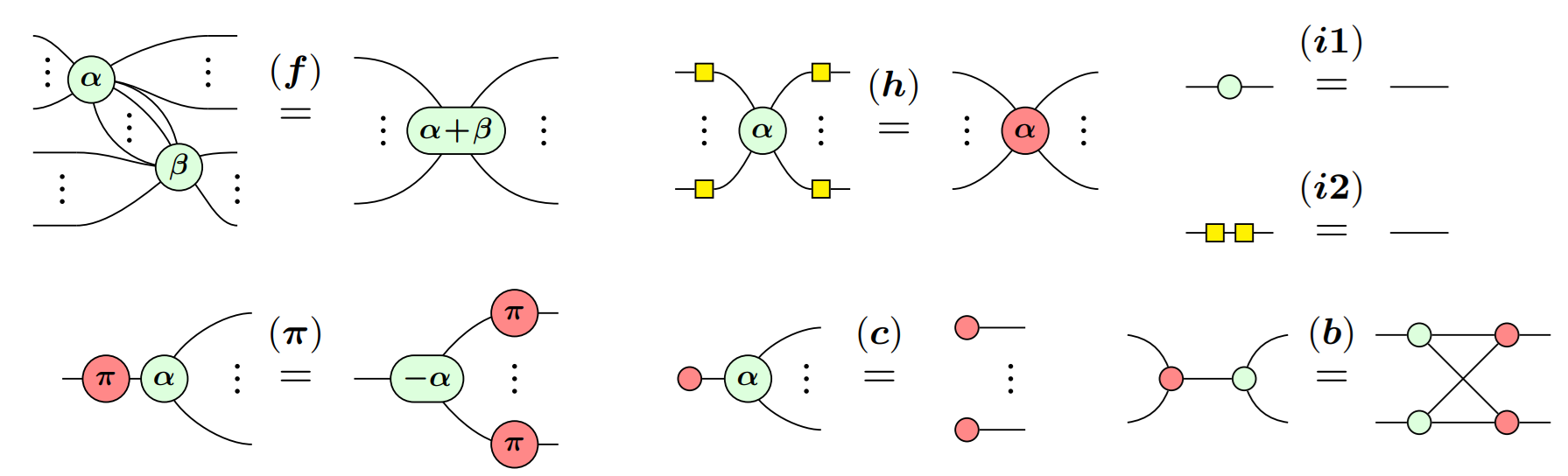}
    \caption{ZX-Calculus Rules. These rules hold for all $\alpha,\beta\in[0,2\pi)$. As a result of rules (h) and (i2), all rules also hold with the colors interchanged. Adapted  from~\cite{Duncan2020graphtheoretic}.}
    \label{fig:zx-rules}
\end{figure}

The ZX-Calculus provides a set of graphical rules for transforming equivalent diagrams. Fig.~\ref{fig:zx-rules} shows a basic set of rules that defines the ZX-Calculus. Additionally, there are important rules that derive from the basic set of rules, such as the Hopf rule, which can be proven through mathematical manipulations or by applying rules from the given set (see Fig.~\ref{fig:antipode-rule}).

\begin{figure}
    \centering
    \includegraphics[width=0.6\linewidth]{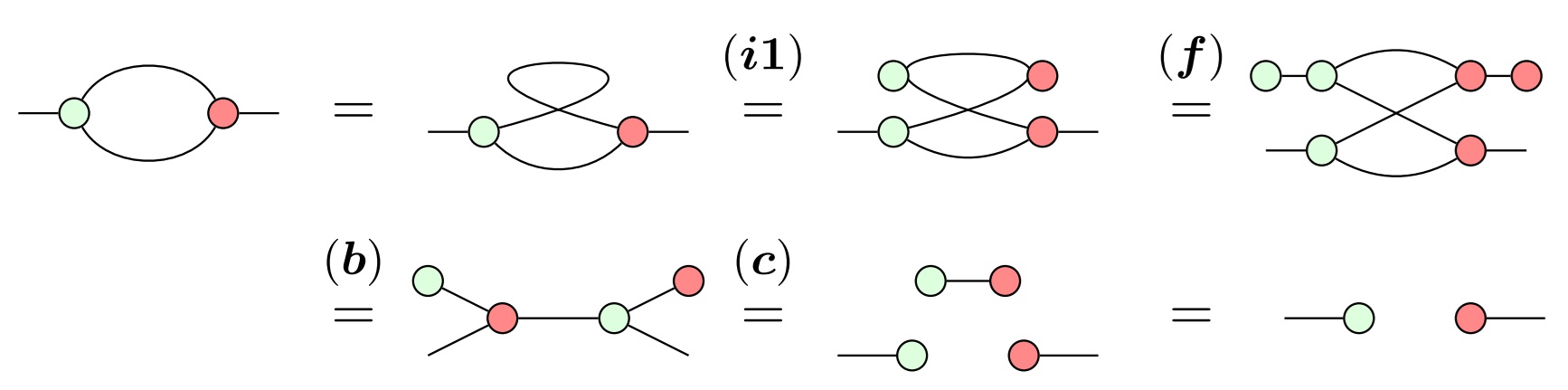}
    \caption{Hopf Rule, derived from the rules in Fig.~\ref{fig:zx-rules}, reveals that even if the nodes seem to be nominally connected, they need not be related at all. Adapted from~\cite{Duncan2020graphtheoretic}.}
    \label{fig:antipode-rule}
\end{figure}

\section{ZX-Calculus Simplification Algorithms}\label{appendix:simp-algo}
To implement the simplification process numerically, we use a Python package, PyZX~\cite{Kissinger_2020}. This package provides a comprehensive platform for creating, visualizing, and simplifying ZX-diagrams. We employ a built-in simplification algorithm designed to bring Clifford circuits to a normal form that is asymptotically optimal in size \cite{Duncan2020graphtheoretic}, which we outline below:

\begin{enumerate}
    \item Convert the circuit to a ``graph-like" standard form (Fig.~\ref{fig:graph-like}).
    \item Remove Clifford spiders by iteratively applying the local complementation rule (LC) and pivoting rules [(P1) and its variations (P2), (P3)] (Fig.~\ref{fig:algorithm-rules}).
    \item Repeat step (2) until the diagram size stabilizes.
\end{enumerate}

The algorithm yields another ``graph-like" circuit that is guaranteed to be simpler than or equivalent to the input circuit, as each iteration removes at least one spider. This simplified circuit is then transformed into a classical network representation (see Fig.~\ref{fig:ZX-to-network}), on which classical percolation analysis is applied.

\begin{figure}[htp!]
    \centering
    \includegraphics[width=0.5\linewidth]{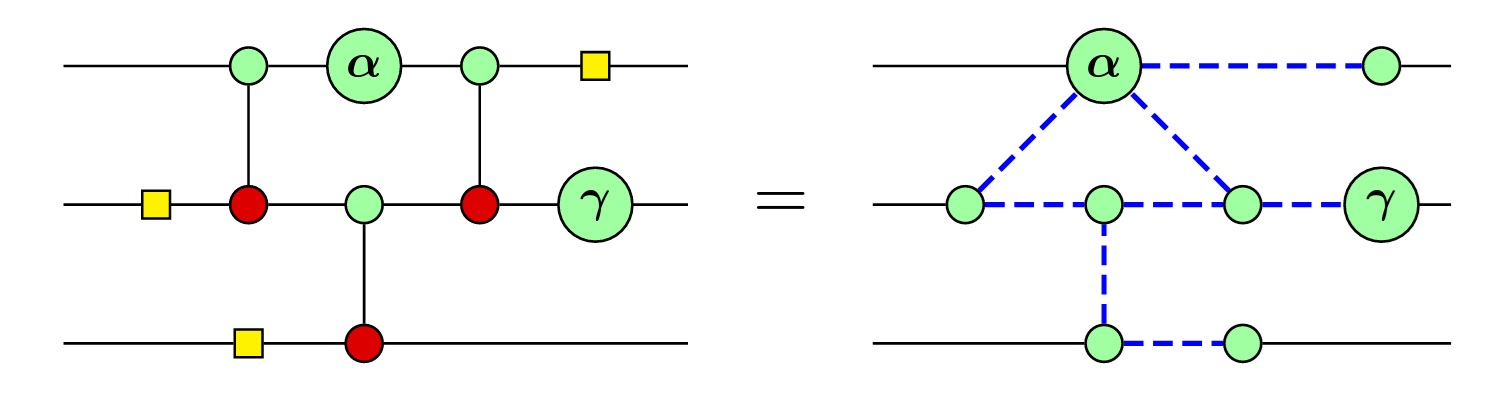}
    \caption{A conversion to ``graph like'' circuits, where (i) all spiders are green, (ii) connected between them only with Hadamard gates (indicated by the dashed blue lines), and (iii) there are no parallel Hadamard gates or self-loops.  Taken from~\cite{Kissinger2020red}}
    \label{fig:graph-like}
\end{figure}

\begin{figure}[htb!]
    \centering
    \includegraphics[width=0.6\linewidth]{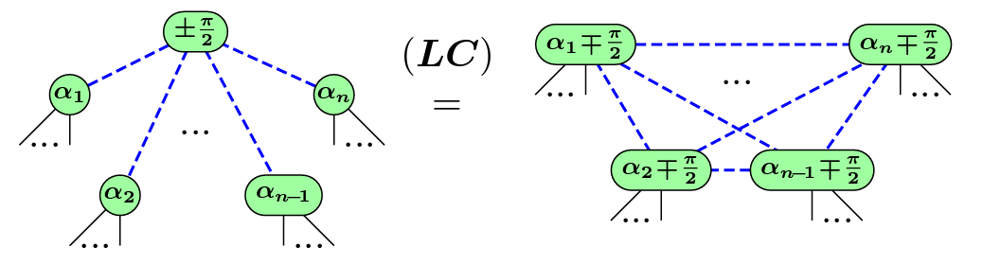}
    \includegraphics[width=0.75\linewidth]{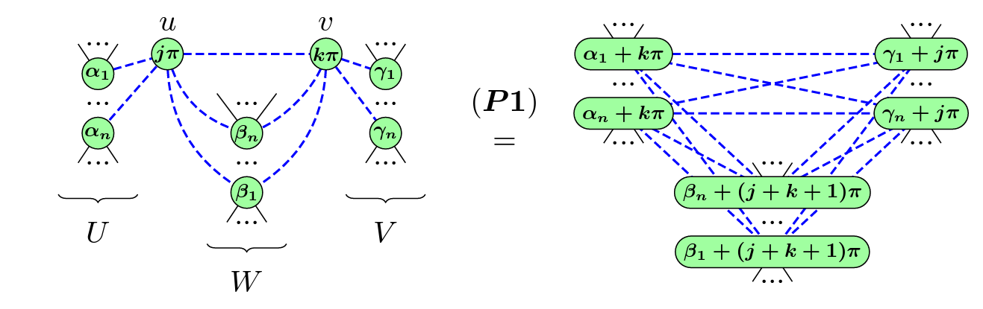}
    \hspace{0.05\textwidth} % Adjust spacing between image and caption if needed
        \caption{Clifford Simplification. (LC) is meant to reduce the $\frac{\pi}{2}$ phase count, and (P1) is meant to reduce the $\pi k$ phase count. Note that they both remove at least one node from the circuit, so their iterations must terminate at some point. Taken from~\cite{Duncan2020graphtheoretic}}
        \label{fig:algorithm-rules}
\end{figure}

\section{Statistical Analysis of the Simplification Process}
One of the main premise of this work is that the Clifford MPT is essentially a classical percolation transition on our ZX simplified networks. However, Refs.~\cite{zabalo2020critical,zabalo2022operator} find deviations between the Clifford CFT and simple two-dimensional bond percolation. Nominally, as long as the operations performed by the ZX simplification are limited to a typical range they can not modify the universal properties at scales much larger than that scale. In this appendix, we investigate the statistical properties of the range of simplification operations performed by our algorithm.

To investigate the spatial extent of the simplification process in monitored Clifford circuits, we analyze the distances over which simplification operations occur. We define a coordinate system for the circuit, where $N$ labels the qubit index and $t$ represents discrete time steps corresponding to circuit layers (the even bond layers are associated with a halved time). The distance between two points in the circuit are defined as $d = \sqrt{(\Delta N)^2 + (\Delta t)^2}$.

The simplification algorithm applies a single rule per step across all applicable locations in the circuit. When two spiders are merged or eliminated, we record the distance between their positions in the $(N,t)$ coordinate system. This approach allows us to quantify the non-locality of the simplification process.

Fig~\ref{fig:sup:statistics}(a) shows the average distance between simplified spiders as a function of algorithm step number. Below the critical measurement probability $p_c$, larger systems exhibit higher average distances, indicating strong system-size dependence. Above $p_c$, this dependence weakens, and distances become more uniform. Fig~\ref{fig:sup:statistics}(b)  presents histograms of simplification distances in the final 25\% of the process, highlighting that simplifications can span distances comparable to the circuit’s maximum extent, especially below $p_c \sim 0.24$. These results provide numerical evidence of the non-local nature of the simplification process, particularly in the volume-law phase. For an intuitive visualization of the simplification process, see the interactive interface ~\cite{Einatscode}.

Finally, in Fig.~\ref{fig:sup:statistics}(c) we plot the same histogram as in Fig.~\ref{fig:sup:statistics}(b), but on a log-log scale and for different values of $N$, where different panels correspond to different values of $p$. The blue solid line corresponds to a distribution $P(d)\sim 1/d$, providing strong evidence that the ZX-simplification process act as a relevant perturbation to the percolation fixed-point, following well known arguments (e.g. Ref.~\cite{FisherMaNickel}). 

\begin{figure}
    \centering
    \includegraphics[width=0.8\linewidth]{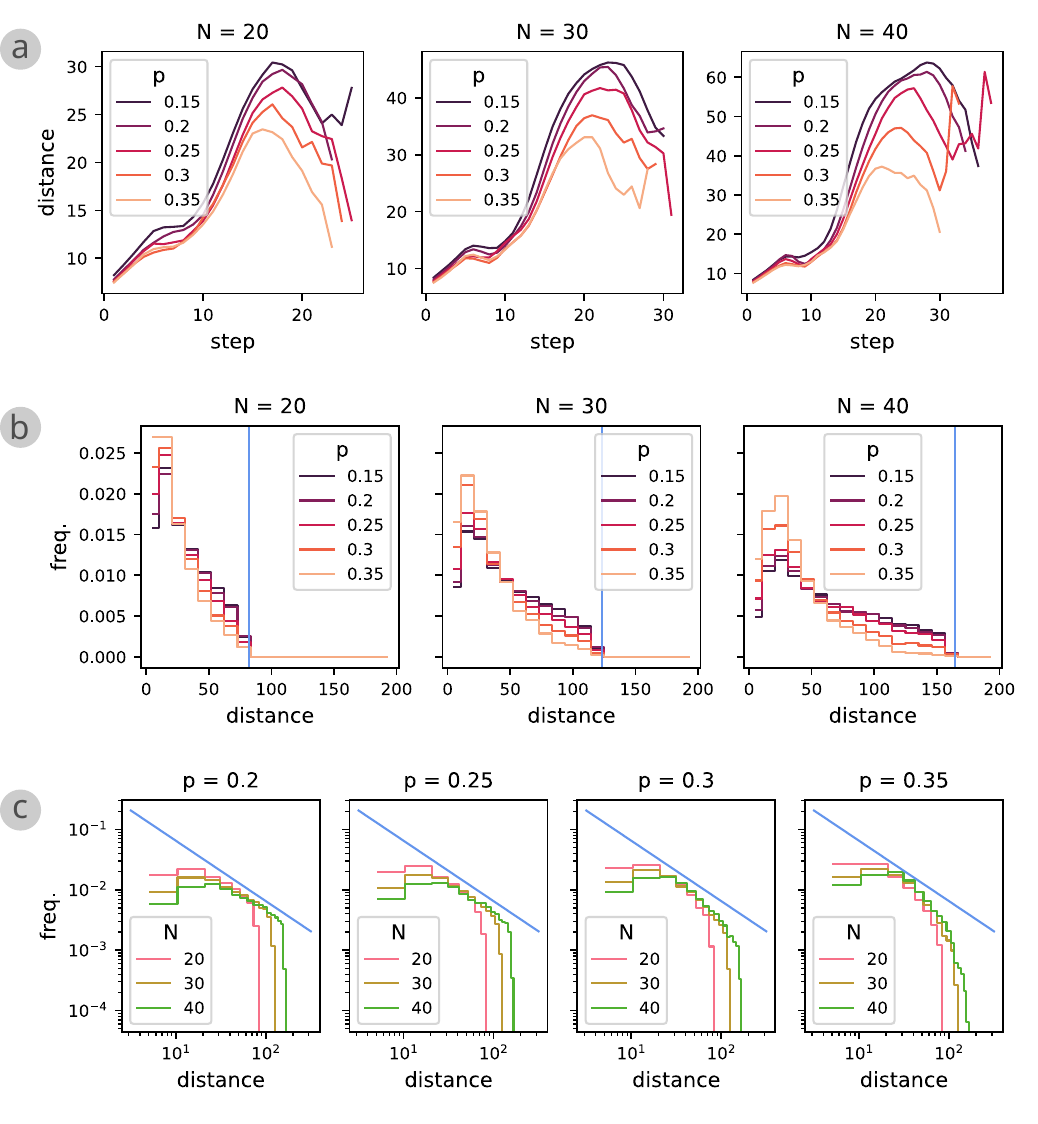}
    \caption{Statistical analysis of simplification distances in monitored Clifford circuits for $r=0.2$ and varying $p$ and $N$, demonstrating the non-local nature of the simplification process. (a) Average distance between spiders undergoing simplification versus algorithm step number. Below the critical point $p_c$, larger systems ($N$) show significantly higher average distances, reflecting strong system-size dependence. Above $p_c$, this dependence diminishes, and the profiles converge. (b-c) Histogram of simplification distances in the final 25\% of the algorithm steps. The blue line in (b) marks the maximum circuit distance, $d_{\text{max}} = \sqrt{N^2 + (4N)^2} = \sqrt{17}N$. Below $p_c = 0.24$, distances extend toward $d_{\text{max}}$ with clear system-size dependence; above $p_c =0.24$, distributions are more localized. (c)  For $p < p_c$, distributions show system-size dependence and span large distances; for $p > p_c$, this dependence diminishes in strength, and the decay with distance is more rapid. The blue line is $1/x$ consistent with a long-ranged distribution close to the critical point.}
    \label{fig:sup:statistics}
\end{figure}

\section{Classical Network Mapping and Percolation Analysis}
We now detail the procedure for analyzing quantum circuit percolation through classical network mapping. Beginning with a simplified graph-like circuit, we construct a classical network where nodes represent both the circuit's spiders and the qubits at initial and final times, while the interconnecting wires form the network's links (see Fig.~\ref{fig:graph-like}). In this representation, percolation occurs when there exists at least one path connecting any input node (initial time qubit) to an output node (final time qubit). Notably, while the original quantum circuits in our model are inherently percolating, the simplification process may yield equivalent circuits that do not maintain this property.

\begin{figure}
    \centering
    \includegraphics[width=0.6\linewidth]{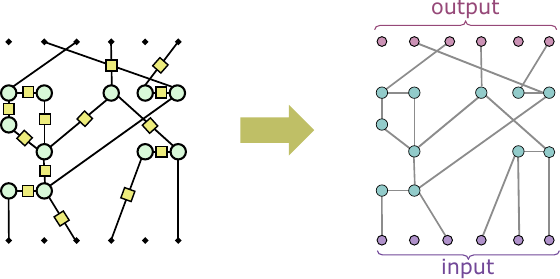}
    \caption{An illustration of the mapping of a graph-like circuit to a classical network.}
    \label{fig:ZX-to-network}
\end{figure}

The percolation probability is computed by averaging over $10^6$ independent realizations. Traditional finite-size scaling analysis typically reveals a universal crossing point in the Binder cumulant curves for different system sizes $N$. However, in our system, we observe a systematic leftward shift in these curves, complicating the direct identification of crossing points. To address this, we employ a phenomenological scaling approach, modeling the percolation probability with a fermionic distribution:
\begin{equation}
f(p)=\frac{1}{e^{(p-p_c)/T}+1}\, ,
\end{equation}
where the effective temperature scales as $T = N^{1/\nu}$ where $\nu=\frac{4}{3}$ is known for classical percolation. This allows us to extract optimal $p_c$ values that exhibit a scaling behavior approximating $p_c=p_{c}(\infty)+{C}/N^{1/\nu}$. The critical point $p_c(\infty)$ is then determined by plotting $p_c$ against $1/N^{1/\nu}$, fitting to a linear function, and extrapolating to $N\to\infty$ (see Fig.~\ref{fig:percolation-examples}). The associated error estimation follows Montgomery et al.~\cite{Montgomery2012}:
\begin{equation}
\mathrm{Var}(p_\infty)=\sigma^2\left(1+\frac{1}{n}+\frac{\bar{x}^2}{\sum_i(x_i-\bar{x})^2}\right)\cdot t_{n-2,\frac{\alpha}{2}}\,;\qquad\sigma^2=\frac{\sum_{i=1}^n{(y_i-\hat{y}_i)^2\cdot w_i^2}}{\sum_i{w_i^2}-2}
\end{equation}
where $x_i=1/\sqrt{N_i}$, $w_i=1/\sigma_i$, $y_i=p_c(N_i)$, $\hat{y}_i$ the value of $p_c(N_i)$ predicted by the fitted line, and $\bar{x}=\sum_{i=1}^n\frac{1}{\sqrt{N}}$. The quantity $\sigma^2$ is the error associated with minimizing the fitting line. Additionally, there is an error of estimating points which are not on this line. This error gets smaller as the number of fitted points is increased, and gets larger as the distance from the points in the original sample grows. The multiplication by $t_{n-\mathrm{dof},\frac{\alpha}{2}}$ gives a factor based on the level of certainty required. The leftward shift of the critical threshold highlights that our algorithm becomes more effective with larger systems, consistent with expectations in the thermodynamic limit.
\begin{figure}
    \centering
    \includegraphics[width=\linewidth]{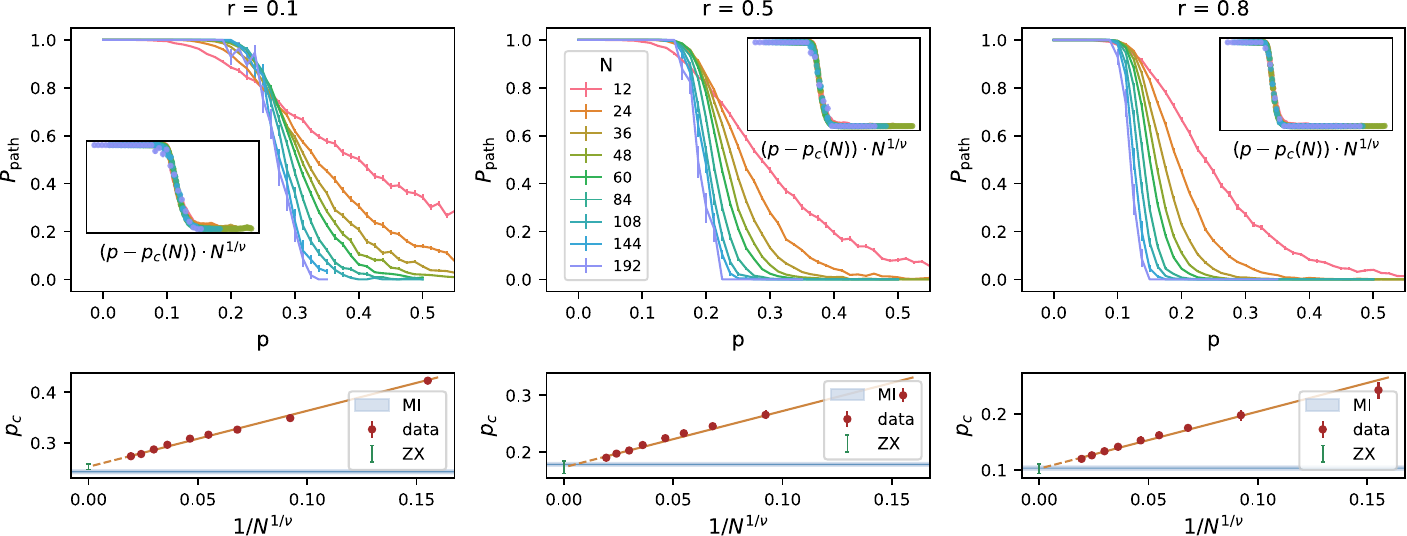}
    \caption{Extended percolation data, with cuts for $r=0.1$ (as in the main text), $r=0.5$ and $r=0.8$}
    \label{fig:percolation-examples}
\end{figure}

\begin{figure}[hbt!]
    \centering
    \includegraphics[width=0.85\linewidth]{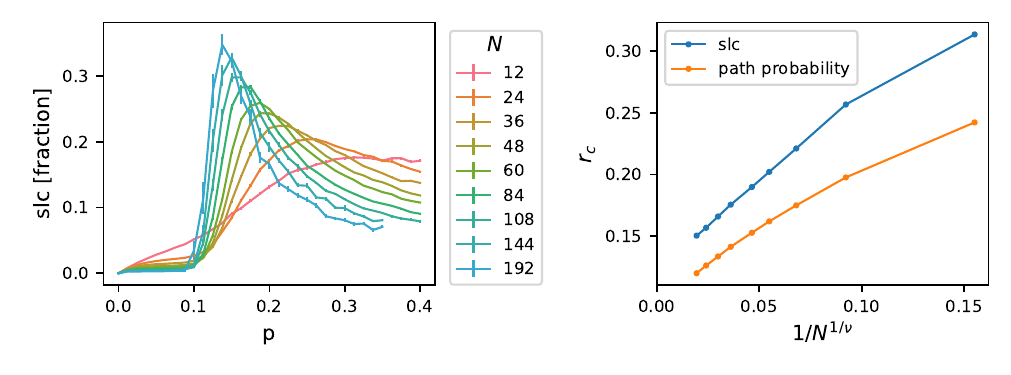}
    \caption{Left: the second largest component's size as a function of $p$ for different system sizes and $r=0.8$. Right: The location of the peak of the second largest cluster as a function of $1/N\rightarrow 0$. }
    \label{fig:slc}
\end{figure}

\section{Analysis of the Second Largest Cluster}

The behavior of the second-largest cluster (SLC) provides an established method for identifying percolation transitions. In conventional percolation theory, the SLC exhibits a characteristic peak near the critical point. This phenomenon can be understood through three distinct regimes:

\begin{enumerate}
    \item Above the percolation threshold: The system consists of multiple small clusters that grow in size as $p$ decreases.
    \item Near the critical point: These clusters reach their maximum size, leading to a peak in the SLC.
    \item Below the critical point: The emergence of a giant component progressively absorbs smaller clusters, causing the SLC to decrease.
\end{enumerate}

We observe a shift in the SLC peak position towards lower $p$ with increasing system size  (see left panel of Fig.\ref{fig:slc}), consistent with the shift in the crossing points of $P_{path}$ (see right panel of Fig.\ref{fig:slc} and Fig.~\ref{fig:findPathProb}).
The peak location shows the same scaling $\sim p_{c}(\infty)+C/N^{3/4}$, although the interception point at $N\to\infty$ is at a higher value.  This deviation is however below our resolution, which is controlled by the width of the peak. The width remains of the order of $0.1$ for the largest system sizes.
\end{widetext}
\end{document}